\documentclass[preprint,showpacs,preprintnumbers,amsmath,amssymb]{revtex4-2}
\usepackage{graphicx}
\usepackage{bm}
\usepackage{color}
\usepackage{xcolor}
\usepackage{caption}
\usepackage{ragged2e}
\usepackage{subcaption}
\usepackage{amssymb}
\usepackage{epsfig}
\usepackage{multirow}
\usepackage[colorlinks=true, allcolors=blue]{hyperref}
\usepackage{array}

\newcolumntype{M}[1]{>{\centering\arraybackslash}m{#1}}

\newcommand{\bk}[1]{\langle #1 \rangle}
\newcommand{\pdy}[1]{\frac{\partial #1}{\partial \Lambda}}

\begin{document}

\title{Entanglement dynamics of many-body quantum states: sensitivity to system conditions and a hidden universality}
\author{Devanshu Shekhar and Pragya Shukla}
\affiliation{ Department of Physics, Indian Institute of Technology, Kharagpur-721302, West Bengal, India }
\date{\today}

\widetext

\begin{abstract}

We consider physical Hamiltonians that can be represented by the multiparametric Gaussian ensembles, theoretically derive the state ensembles for its eigenstates and analyze the effect of varying system conditions on its bipartite entanglement entropy. Our approach leads to a single parametric based common mathematical formulation for the evolution of the entanglement statistics of different states of a given Hamiltonian or different Hamiltonians subjected to same symmetry constraints. The parameter turns out to be a single functional of the system parameters  and thereby reveals  a deep web of  connection hidden underneath different quantum states.

\end{abstract}

\maketitle

\section{Introduction} \label{intro}

An important requirement of quantum information processing is 
that any knowledge of the many-body state provides minimum information  about its subunits. One way to  ensure this is by maximizing the amount of entanglement among the subunits; this in turn  leads to an extensive search for the states with maximum entanglement, e.g., an ergodic  state (i.e. one accessing all parts of Hilbert space with equal probability).  An arbitrary many-body state is however usually non-ergodic with partial entanglement (lying between separability and maximum entanglement).   
The question is
(i) how to quantify its entanglement?
(ii) what type of variation of system conditions can lead to an enhancement of entanglement?
(ii) whether its entanglement can be preserved in time or under changing system conditions once it reaches the maximum entanglement limit?
Notwithstanding intense research efforts in past two decades, the answers to many such questions  are still not available. The growth of entanglement among subunits with time also renders the analysis beyond the access of numerical techniques. This motivates us to seek new theoretical routes to investigate many-body entanglement and thereby gain insights in the non-equilibrium quantum dynamics.

In general, the interactions among various subunits of a many-body system are sensitive to a host of system conditions, both static as well dynamic type. A variation of these conditions can lead to changes in the mutual interactions among subunits and is expected to manifest, thereby, in the dynamics of entanglement measures too, more clearly if the dynamics is analyzed in the non-interacting basis, i.e., product states of subunits.  Indeed, the ideally sought information in context of many-body systems  is the one that describes multi-partite entanglement for a fixed set of system conditions as well its variation with changing interactions among various parts; the technical complexity however renders the determination a very difficult task. Fortunately many important insights can still be achieved by consideration of the entanglement between its two sub-parts, referred as the bipartite entanglement and is the main objective of the present study.

The standard route to determine the entanglement measures for a pure state requires a prior knowledge of the density matrix $\rho =  |\Psi\rangle \langle \Psi|$. The latter can in turn be obtained  by solving the eigenvalue equation for the Hamiltonian, determining its states and then calculation of the state matrix. An alternative route adopted in many previous studies was either based on direct modelling of the state matrix ensemble, e.g., by a Ginibre ensemble leading to a stationary Wishart ensemble (e.g. \cite{zyczkowski2011generating,nadal2011statistical,kumar2011entanglement,majumdar2010extreme,Collins2016,bengtsson2017geometry}) or its generalizations, e.g., \cite{psds1, psds2, psds3}. Although these studies provided many insights, they lacked a very important aspect: as the state matrix ensemble was derived  without any reference to underlying Hamiltonian, the connection of ensemble parameters with system parameters was not explicit. This handicaps one from a direct analysis of the entanglement dynamics with changing system parameters. The primary objective of our present study is to fulfil this information gap.

To achieve our objective, we first need to determine the matrix representation of the Hamiltonian in a physically motivated basis and thereafter its eigenstates. A many-body Hamiltonian however consists of complicated interactions among its subunits which even if well-known, e.g., coulomb interactions, an exact determination of its matrix elements is not often possible; this could occur, for example, due to technical issues encountered in calculation of the integrals either by theoretical or numerical route. The incomplete knowledge or error in their determination manifests itself by randomization of the matrix elements. Indeed, the nature and type of the distribution of the matrix elements is sensitive to various system conditions, e.g., symmetry and conservation laws, dimensionality and boundary conditions, disorder etc. and can vary from one element to the other.
As a consequence, the  Hamiltonian matrix is best represented  be a system-dependent random matrix, with some or all elements randomly distributed, the type and the strength of the latter in general sensitive to underlying system conditions.

As expected, based on eigenvalue equation,  the randomness underlying  the matrix elements  manifests in the eigenstates of the Hamiltonian  too and the distribution of the latter can be derived, in principle, from the JPDF of the former by a transformation of variables from matrix space to eigenvalue-eigenstate space.  An integration of the ensemble density of the Hamiltonian, i.e, the JPDF  of the matrix elements is in general technically complicated, thus motivating a search for alternative routes, e.g., a differential route. As discussed in \cite{pseig}, an evolution equation for the JPDF of the components of an arbitrary eigenstate (referred as state JPDF hereafter for brevity) for a many-body system represented by multiparametric Gaussian ensembles can be derived from the ensemble density. The information can further be used to determine  the statistical behavior of the entanglement measures. An additional benefit of the differential route is a common mathematical formulation in terms,  of the state JPDF for a wide range of many-body systems where the system information is contained in a single parameter, i.e., the complexity parameter. We pursue the above  approach in the present work to derive the evolution equation for the entanglement measures and their solutions.

The paper is organized as follows. We begin, in section \ref{state}, with two examples of prototypical Hamiltonians: (i) a many-body Hamiltonian modelled by a generalized version of the quantum random energy model (QREM) and (ii) the random-field Heisenberg model (RFHM), their ensemble densities, and the relation between system parameters and ensemble parameters. These examples help to analyze the effect of changing system conditions on the ensemble parameters. Section \ref{density} describes briefly necessary tools for our approach, i.e, the diffusion equation for the ensemble density of a many-body Hamiltonian in terms of changing  ensemble parameters. This is necessary to derive the matrix elements moments used, in section \ref{diff_density}, to derive the diffusion equation for the components of an arbitrary many-body state. (An alternative route to derive  the equation, based on a direct integration of the Hamiltonian ensemble density,  is discussed in \cite{pseig}).  Section \ref{schmidtEvol} presents the derivation evolution equation of the Schmidt eigenvalues, which is then used in the Section \ref{evolmeas} in the calculation of the average and variance of an entanglement measure, the von Neumann entropy, for bipartite pure states with varying system conditions.
  As the theoretical derivations are technically involved, we have moved  them to supplementary material \cite{sup} to avoid distraction from the significance of our results. Based on the two systems mentioned in section \ref{state}, we also pursue a numerical verification of our theoretical predictions and find a good agreement; this is described in section \ref{numerical}. We conclude in section \ref{conclusion} with a summary of our ideas, results and open questions.

\section{Definitions: state matrix and entanglement entropies} \label{state}

Consider a many-body system in a pure state $|\Psi \rangle$. The information about the quantum correlations between the two sub-parts is contained in the components of $|\Psi \rangle$ represented in the bipartite basis, consisting of orthonormal bases of the two sub parts. Assuming the orthogonal subspaces of its sub parts $A$ and $B$, consisting of basis vectors $|a_k \rangle$ and $| b_l \rangle$, $k=1 \to N_A, l=1 \to N_B$ respectively, we can write $|\Psi \rangle = \sum C_{kl} \; | a_k \rangle \; | b_l \rangle$; the matrix $C$ with coefficients $C_{kl}$ (complex or real) as its entries is referred as the state matrix. 

The standard entanglement measures for a pure bipartite state, viz., the von Neumann entropy $R_1$ and  R\'enyi entropies $R_n$ with $n >1$, are functions of the eigenvalues  of  the $N_A \times N_A$ reduced density matrix $\rho_A$ for subsystem $A$ where  $\rho_A = CC^{\dagger}$. The eigenvalues of $\rho_A$ are also referred  as the Schmidt eigenvalues. Denoting the latter as $\lambda_1, \ldots, \lambda_N$, for example, the  von Neumann entropy is defined as $R_1 = - {\rm Tr} \rho_A \; {\rm log} \rho_A = - \sum_{n=1}^{N_A} \lambda_n \; {\rm log} \lambda_n$ and $R_{\alpha} = {1\over 1-\alpha} {\rm ln} \; {\rm Tr} (\rho_A)^{\alpha}=  {1\over 1-\alpha} {\rm ln} \; \sum_{n=1}^{N_A} \lambda_n^{\alpha}$.
  
\section{Hamiltonian and the representative ensemble} \label{density}

A determination of the Schmidt eigenvalues requires prior information about the state  matrix $C$, i.e., the components of $|\Psi \rangle$ in a bipartite basis. The latter can in principle be obtained by solving the eigenvalue equation $H \, |\Psi\rangle = \lambda \, |\Psi\rangle$ with $H$. This in turn requires the determination of the matrix elements of the many-body Hamiltonian in the basis. But, as mentioned in section \ref{intro}, the error associated with exact determination of the matrix elements, further complicated by the fluctuating system conditions, render it necessary to describe the matrix elements by a statistical distribution in a physically motivated basis, e.g., the bipartite basis in the present case. Due to technical complexity of our theoretical ideas, it is helpful to first consider some prototypical examples.

{\bf (i) Quantum random energy model (QREM):} 
We consider a one dimensional lattice of $L$ spin $1/2$ particles in a random magnetic field, described by the Hamiltonian
\begin{equation}
    H = H_{REM} + \Gamma \sum_{i=1}^L \sigma_i^x, 
\label{qrem}
\end{equation}
where, $H_{REM} = \sum_k E_k |k \rangle \langle k|$ is the random energy part of the Hamiltonian, diagonal in the $S_z$ basis $\{|k\rangle\}$, such that the energies $\{E_k\}$ are independent and identically distributed as
$ P(E_k; h, L) = \frac{1}{\sqrt{\pi L}} \, e^{-\frac{E_k^2}{ L}}$ and $\sigma_x$ is one of the Pauli matrices. For $\Gamma$ as a (negative) constant, eq. \eqref{qrem} corresponds to the Hamiltonian of  the well studied QREM \cite{Laumann2014,Baldwin2016,Biroli2021,Parolini2020} that can be mapped to an Anderson model on an $L$-dimensional hypercube with the spin configurations being lattice sites with a constant hopping strength \cite{Baldwin2016}. For a fixed energy density e.g. $\epsilon = 0$, the dynamics of QREM is governed by the transverse-field  parameter  $\Gamma$. In the matrix space, however, this leads to a mere translation of the off-diagonal elements and there is an absence of diffusion. This is the case in general when the variance of all the matrix elements are kept fixed.

For a matrix representation of $H$, we choose the product state basis spanned by $2^L$ vectors of type  $|\mu \rangle \equiv \prod_{k=1}^L |m_k \rangle$ with each $\sigma_z|m_k\rangle = \pm |m_k\rangle$. A state $\Psi$ of $H$ in this basis can be expressed as $|\Psi\rangle \equiv \sum_{\mu} \psi_{\mu} |\mu \rangle $. For the entanglement analysis, we consider a bi-partition of the spin-chain into two equal subparts, referring them as subsystem  $A$ and $B$ with their local basis spaces as $|a_{\mu}\rangle$ and $|b_{\mu} \rangle$. The basis state $|\mu \rangle$ can then be written as the $|\mu \rangle \equiv |a_{\mu} \, b_{\mu}\rangle$. The components of $|\Psi\rangle$ can now be written $C_{a_{\mu} \, b_{\mu}}$: $\psi_{\mu}  \equiv  C_{a_{\mu} \, b_{\mu}} \equiv \langle a_{\mu} \, b_{\mu}|\Psi\rangle$. In the $\sigma_z$ basis, the non-zero off-diagonal elements correspond to the basis states at a Hamming distance one from the diagonal; (this is because $\sigma_i^x$ flips the spin at only one site $i$). This leads to $L$ non-zero off-diagonal elements per row in the matrix representation of the Hamiltonian given by eq. \eqref{qrem}.

In order to analyze the multiple system parameters dependence for this model, we consider $\Gamma$ to be a basis-dependent random variable  with mean zero and a power-law decaying variance i.e. $\langle \Gamma_{\mu \nu}^2 \rangle = \frac{1}{1 + (\frac{\mu-\nu}{b})^2}$, $\langle \Gamma_{\mu \nu} \rangle = 0$, which can then be mapped to an Anderson model on an $L$-dimensional hypercube with a random and anisotropic hopping strength. This in turn leads to independent Gaussian distributed off-diagonals with  mean $\langle H_{\mu \nu} \rangle = 0$ and variance $\langle H_{\mu \nu}^2 \rangle = \frac{1}{1 + (\frac{f_{\mu \nu}}{b})^2}$ where $f_{\mu \nu} = |\mu-\nu|^2$ if the basis elements $\mu$ and $\nu$ are at a unit Hamming distance from each other, $\langle H_{\mu \nu}^2 \rangle=0$ otherwise.  With diagonal $H_{kk} = E_k$, the diagonals $H_{kk}$ are distributed as Gaussians with zero mean and same variance: $\rho(H_{\mu \mu}) \propto \exp\left(-\frac{H_{\mu \mu}^2}{L}\right)$. 
The ensemble density i.e. the JPDF of the matrix elements of eq. \eqref{qrem} can now be written as
\begin{equation}
    \rho_1(H) \propto \prod_{\mu} \exp\left(-\frac{H_{\mu \mu}^2}{L}\right) \prod_{\mu < \nu} \; \exp\left[-\left(1 + \frac{f_{\mu \nu}}{b^2}\right)\frac{H_{\mu \nu}^2}{2}\right],
    \label{rho1}
\end{equation}
where, $\mu, \nu$ correspond to only connected pairs. For later reference, we rewrite eq.(\ref{rho1})  as 
\begin{eqnarray}
\rho(H) = \mathcal{N} \;  {\rm exp}\left[-  \sum_{\mu, \nu} \;{1\over 2 v_{\mu\nu}} \left( H_{\mu\nu} -b_{\mu\nu} \right)^2 \right] 
\label{rhog}
\end{eqnarray}
with  $v_{\mu\nu}=\langle (H_{\mu\nu} )^2\rangle-\langle H_{\mu\nu} \rangle^2$ and  $b_{\mu\nu} =\langle H_{\mu\nu} \rangle$ and $\mathcal{N}$ is the normalization constant.  A comparison of eq.(\ref{rho1}) with eq.(\ref{rhog}) gives 
\begin{equation}
v_{\mu \mu}=L/2,  \hspace{0.1in}  v_{\mu\nu} = \left(1 + \frac{f_{\mu \nu}}{b^2}\right)^{-1}, \hspace{0.1in} b_{\mu\nu} = 0.
\label{qrem2}
\end{equation}
With vanishing variance of the off-diagonal elements for $b = 0$, the system becomes classical (Poissonian) (as in the case for non-random $\Gamma = 0$). For $b \to \infty$ the system goes to the GOE limit (analogous to  large non-random $\Gamma$ case).

\vspace{0.2in}

{\bf (ii)  Random-field Heisenberg model (RFHM):}  The Hamiltonian in this case describes the dynamics  of $L$ spin $1/2$ particles on a one dimensional lattice in a random magnetic field, 
\begin{equation}
    H = \sum_{k=1}^{L} \left[J \left(S_k^x S_{k+1}^x + S_k^y S_{k+1}^y\right) + D \, S_k^z S_{k+1}^z - h_k S_k^z\right], 
\label{xxz}
\end{equation}
where $2 S^x, 2 S^y, 2 S^z$ are Pauli matrices,  $J$ is a constant which sets the energy scale, $D$ is the anisotropy parameter, and the fields  $h_i$ are Gaussian random variables. The Hamiltonian conserves the total magnetization $S_z^{tot} = \sum_{i=1}^L S_i^z$ even in presence of the disorder \cite{luitz2015many,luitz-dist,buijsman2019random}. 

For a matrix representation of $H$, we choose the product state basis spanned by $N=2^L$ vectors of type  $|\mu \rangle \equiv \prod_{k=1}^L |\mu_k \rangle$ with each $S_k|\mu_k\rangle = \mu_k |\mu_k\rangle$ with $\mu_k=\pm 1/2$. A state $\Psi$ of $H$ in this basis can be expressed as $|\Psi\rangle \equiv \psi_{\mu} |\mu \rangle $. For the entanglement analysis, we consider a bipartition of the spin-chain into two equal subparts, referring them as subsystem  $A$ and $B$ with their local basis spaces as $|a_{\mu}\rangle$ and $b_{\mu} \rangle$. The basis state $|\mu \rangle$ can then be written as the $|\mu \rangle \equiv |a_{\mu} \, b_{\mu}\rangle$. The components of $|\Psi\rangle$ can now be written $C_{a_{\mu} \, b_{\mu}}$: $\psi_{\mu}  \equiv  C_{a_{\mu} \, b_{\mu}} \equiv \langle a_{\mu} \, b_{\mu}|\Psi\rangle$.

With  onsite disorder,  the diagonal elements of $H$ are  correlated  random variables in the chosen basis. The nonzero off-diagonals are however non-random, equal to $J/2$. The JPDF of the matrix elements of the Hamiltonian in eq.(\ref{xxz}) can then be written  as

\begin{equation}
    \rho_1(H) = C_1 \; \exp\left[- {\mathcal H}_D^{T} \, B_D \, {\mathcal H}_D  \right] \prod_{\mu < \nu} \; \delta \left(H_{\mu\nu}-\frac{J}{2}\right),
    \label{jpdfxxz0}
\end{equation}
with $C_1$ as the normalization constant, ${\mathcal H}_D$ as a column vector of size $N$, consisting of the diagonals $H_{\mu \mu}-\langle H_{\mu \mu} \rangle$ with $H_{\mu \mu} = \equiv \langle \mu | H | \mu \rangle$ of the Hamiltonian $H$ in eq.(\ref{xxz}), and $B_D$ as the precision matrix, $B_D=\Sigma_D^{-1}$, where $\Sigma_D$ is the $N \times N$ covariance matrix of the elements in ${\mathcal H}_D$: 
$\Sigma_{D;\mu\nu } = \langle H_{\mu \mu} H_{\nu \nu} \rangle - \langle H_{\mu \mu} \rangle \langle H_{\nu \nu} \rangle = \sum_{k=1}^L \mu_k \nu_k \langle h_k^2 \rangle$ (details discussed in the supplementary file \cite{sup}). Assuming  $\langle h_k^2 \rangle = v^2$, we then have  $\Sigma_{D;\mu \nu} = \beta_{\mu \nu} \, v^2$ with $\beta_{\mu \nu} =\sum_{k=1}^L \mu_k \nu_k$ and $v^2 = L h^2 / 4$.   Again replacing $\delta$-functions by limiting Gaussians, the above $\rho_1(H)$ can be rewritten as  $\rho_1(H)  = \lim_{\Sigma_{\mu\nu, \mu' \nu'} \to 0} \; \rho(H)$ where

\begin{eqnarray}
\rho(H) &=& C_1 \; \exp\left[- {\mathcal H}^{T} \, B \, {\mathcal H}  \right] \\
&=& C_2 \;  {\rm exp}\left[-  \sum_{\mu \nu; \mu', \nu'} B_{\mu \nu; \mu' \nu'}\; H_{\mu \nu} H_{\mu' \nu'} -A_{\mu \nu} H_{\mu \nu} \right]
\label{rhoco1}
\end{eqnarray}
with ${\mathcal H}$ as a column vector of size $M=N(N+1)/2$, consisting of all  diagonals and upper off-diagonals elements $H_{\mu \nu}- \langle H_{\mu \nu}\rangle$ and $B=\Sigma^{-1}$ where $\Sigma$ is the new covariance matrix of size $M \times M$: $\Sigma_{\mu \nu, \mu' \nu'} = \langle H_{\mu \nu} H_{\mu' \nu'} \rangle - \langle H_{\mu \nu} \rangle \langle H_{\mu' \nu'} \rangle$. Here $A_{\mu \nu}= -2 \sum_{\mu' \nu'} B_{\mu \nu, \mu' \nu'} \langle H_{\mu' \nu'}\rangle$. As discussed in detail in the supplementary file \cite{sup}, we have 
\begin{eqnarray}
B_{\mu\mu, \nu \nu} &=& \left(\frac{1}{v^2}\right) \, \beta^{-1}_{\mu \nu}, \hspace{0.2in}
B_{\mu \nu, \mu \nu} =0  \qquad {\mu \not= \nu}, \nonumber \\
A_{\mu\nu} &=& -\left({2 D \over v^2}\right) \, \alpha_{\nu} \, \beta^{-1}_{\mu \nu}, \hspace{0.2in} 
 \langle H_{\mu\nu}\rangle = D \, \alpha_{\mu} \, \delta_{\mu \nu} + {J\over 2} \, (1-\delta_{\mu \nu})
\label{jpdfxxz1}
\end{eqnarray}

The system, eq. \eqref{xxz}, exhibits an ergodic phase (delocalized wavefunctions) for low disorder and a localized phase as the disorder strength becomes large.  This is consistent with  its accepted ensemble representation: $\rho(H)$ corresponds to a Gaussian orthogonal ensemble (GOE) in  low disorder limit which has delocalized eigenfunctions,   and,  a Poisson ensemble in strong disorder limit with localized eigenfunctions.

\section{Variation of System Parameters: response of Hamiltonian ensemble density} \label{diff_density}

In contrast to the eigenvalues, the eigenstates of an operator are basis dependent. With primary focus on the eigenstates dynamics in this work, it is relevant to choose an appropriate basis for the Hamiltonian matrix representation. In case a Hamiltonian has discrete global symmetries, an appropriate basis for its representation is a symmetry resolved basis. The latter leads to a block diagonal structure and ensures that each block has non-degenerate eigenvalues. With physical properties of different blocks uncorrelated, it is then sufficient to consider an ensemble of symmetry resolved blocks, with different members of the ensemble sharing the same set of global constraints (e.g., quantum numbers, symmetry conditions  and conservation laws).   For clear exposition of our ideas without loss of generality, here we consider a Hamiltonian with non-degenerate eigenvalues. 

The ensemble density $\rho$ in eq.(\ref{rhog}) and eq.(\ref{rhoco1}) is a function of matrix elements $H_{kl}$ as well as ensemble parameters and thereby a functional of system parameters through both of them. 
A change in system conditions can therefore change  $\rho$ through matrix elements, leading to its dynamics in matrix space, or through a change of matrix elements moments, leading to the dynamics of $\rho(H)$ in $\{v,b \}$ space. As the dynamics in the two spaces (i.e. matrix elements and ensemble parameters)  is of the same system due to same cause, both of them are expected to be related.  As discussed in a series of previous studies  \cite{psflat, psand, psco, psalt}, this is indeed the case: the evolution of the ensemble densities $\rho(H)$ given by eq.(\ref{rhog}) or eq.(\ref{rhoco1}) with changing ensemble parameters and their complexity parameter formulation is discussed in detail in \cite{psalt, psco, pseig}. The previous analysis also indicated the existence of a universal ensemble representation for a wide range of  systems e.g. those described by eq.(\ref{qrem}) or eq.(\ref{xxz}); the ensemble evolves with respect to a single parameter only that is a function of all system parameters.

A change of  $\rho$ is expected to  manifest on the eigenfunction components in the basis space and their statistical behavior. To avoid confusion with state ensemble used in this work,   hereafter we refer $\rho(H)$ as the Hamiltonian ensemble. For purpose of clarity, here we briefly review the complexity parameter formulation for the ensemble described by eq.(\ref{rhog}).

As discussed in previous studies \cite{psflat, psand, psco, psalt}, a specific combination $T \rho \equiv \sum_{\mu \leq \nu} [(g_{\mu \nu} - 2 \gamma v_{\mu \nu}) \frac{\partial \rho}{\partial v_{\mu \nu}} - \gamma b_{\mu \nu} \frac{\partial \rho}{\partial b_{\mu \nu}}]$ of first order variation of the ensemble parameters $v_{\mu \nu;s} \to v_{\mu \nu;s} + \delta v_{\mu \nu;s}$ and $b_{\mu \nu;s} \to b_{\mu \nu;s} + \delta b_{\mu \nu;s}$  over time would lead to a Brownian dynamics  in Hermitian matrix space, starting from an arbitrary initial condition and with a stationary ensemble as the equilibrium limit: $T \rho =L \rho$ where $L \equiv \sum_{\mu, \nu}{\partial \over \partial H_{\mu \nu;s}}\left[{g_{\mu \nu}\over 2}{\partial \over \partial H_{\mu \nu;s}} +  \gamma H_{\mu \nu; s}\; \right]$ with $g_{\mu \nu}=2$ or $1$ for $\mu =\nu$ and $\mu \not=\nu$, respectively.  A transformation of  the set $\{ v_{\mu \nu}, b_{\mu \nu} \}$ to another set $\{t_1, t_2, \ldots, t_M \}$ however reduces the multi-parametric dynamics in the ensemble space to a single parameter dynamics, say with respect to $t_1$, while others, i.e, $t_2 \ldots t_M$ remaining constant throughout the evolution,
\begin{eqnarray}
    {\partial \rho\over\partial t_1} &=&  L \, \rho,
\label{rhot}
\end{eqnarray}
where,
The above condition can be fulfilled by three possible ways (details discussed in supplemental material \cite{sup}):

\begin{eqnarray}
{\rm {Case \; I}} \qquad  && T t_1=1, \; T t_n=0 \; \;  n >1, \qquad \frac{\partial \rho}{\partial t_{\alpha}} = 0 \quad \forall \; \alpha >1,  
\label{yc1}\\
{\rm {Case \; II}} \qquad  && T t_1=1, \; \frac{\partial \rho}{\partial t_{\alpha}} = 0 \quad \forall \; \alpha >1 \hspace{1.5in} \label{yc2}\\
{\rm {Case \; III}} \qquad && T t_1=1, \; T t_n=0  \qquad \forall \; \;  n >1,  \hspace{1.5in} 
\label{yc3}
\end{eqnarray} 
The parameters $t_1, \ldots, t_M$ for each one of the above cases can be obtained  by solving the characteristic set of equations ${{\rm d}y_{kl;s} \over A_{kl;s}}
= {{\rm d}x_{kl;s} \over B_{kl;s}}={{\rm d}t_{\alpha} \over \delta_{\alpha 1}}$ with $k,l=1 \to N$. Here $M$ corresponds to total number of ensemble parameters participating in evolution. Here $M=2 N^2$ for all of them varying, $M < 2 N^2$ in case ${\partial \rho_1 \over \partial y_{kl}}=0$ or ${\partial \rho_1 \over \partial x_{kl}}=0$ for some $y_{kl}$ or $x_{kl}$, with $k,l$ arbitrary. For example,
for the case with $x_{kl} = 0$ ($\forall k,l$), we have $M = N^2$.

The transformation  maps the JPDF  $\rho$ in eq.(\ref{rhog})   to  $\rho(H; t_1, t_2, \ldots, t_{M})$ with $t_1, t_2, \ldots, t_{M}$ given by a set of characteristic equations  \cite{psalt}
\begin{eqnarray}
\frac{d v_{\mu \mu;s}}{f_{\mu \mu;s}} = \ldots = \frac{d v_{\mu \nu;s}}{f_{\mu \nu;s}} = \frac{db_{\mu \nu;s}}{b_{\mu \nu;s}} = \frac{d t_{\alpha}}{\delta_{\alpha 1}}
\label{cheq1}
\end{eqnarray}
with $f_{\mu \nu;s} \equiv (g_{\mu \nu}- 2 \gamma v_{\mu \nu;s})$. A general solution of the above equation for $t_1$ can be given as 
\begin{eqnarray}  
 t_1 = -{1\over  2 M \gamma}  \; \sum_{s=1}^{\beta} \sum_{\mu \le \nu} \; \bigg(q_{\mu \nu} \; {\rm ln} |g_{\mu \nu}-2 \gamma v_{\mu \nu;s}|  + p_{\mu \nu} \; \; {\rm ln}  |b_{\mu \nu;s }|^2 \bigg) + const.
 \label{yparam1}
\end{eqnarray}
with $q_{\mu \nu}$ and $p_{\mu \nu}$ as arbitrary constants dependent on initial conditions, and $\beta$ is the Dyson's index, which is $1 \, (2)$ for real-symmetric (complex-Hermitian) matrices. Choosing $ q_{\mu \nu} =1, p_{\mu \nu} =1$, we have 

\begin{eqnarray}  
 t_1= -{1\over  2 M \gamma}  \; \; {\rm ln}\left[ \prod_{s=1}^{\beta} \prod_{\mu \le \nu}|g_{\mu \nu}-2 \gamma v_{\mu \nu;s}| \, |b_{\mu \nu;s }|^2 \right] + const.
 \label{yparam}
\end{eqnarray}
Similarly $t_2, \ldots, t_{M}$ can be  obtained by solving the eq.(\ref{cheq1}) for $\alpha >1$. As their explicit forms are not needed for 
further analysis, we omit the related details here. Also, we note that $t_2, \ldots, t_{M}$ remain not only constants of the evolution described by eq.(\ref{rhot}), they can also be chosen as the basis constants \cite{psds2}.  
The product in eq.(\ref{yparam}) is over $M$ non-zero terms. As clear from the above, $Y$ turns out to be an average distribution parameter, a measure of average uncertainty of system, also referred as the \textit{ensemble complexity parameter}.  

The transformation of variables $v, b \to t$ maps  the ensemble density $\rho(H, v, b)$ of $H$-matrices to another ensemble density $\rho(H, t)$ (with matrices $H$ remaining same). Equivalently the evolution of the ensemble density $\rho(H)$ due to changing ensemble parameter matrices $v, b$ is mapped to another $M$-dimensional parameter space ``t" consisting of  $t_1, t_2, \ldots, t_M$  as variables and referred as ``complexity space" (to distinguish it from original ensemble parameter space). The evolution in complexity space however occurs along the curves along which  only $Y \equiv t_1$  varies, with $t_2, \ldots, t_M$ remaining constants. These constants  can therefore be determined from the initial condition on  $\rho(H)$. As discussed later, the above in turn leads to the evolution equations   for the state matrix elements of a typical eigenstate of $H$ and thereby for the Schmidt eigenvalues in the complexity parameter space (i.e. Eq.(\ref{pu}) and  Eq.(\ref{pdl0}) respectively).

The complexity parameter formulation for eq.(\ref{rhoco1}), i.e., ensemble densities with pairwise matrix elements correlations can be derived by following the similar steps as mentioned above. The evolution equation for $\rho(H)$ again turns out to be the same form as eq.(\ref{rhot}) but the parameters $t_1, \ldots, t_M$ are now different and are determined by following set of characteristics equations
\begin{eqnarray}
\frac{d B_{\mu \nu, \mu' \nu'}}{ f_{\mu \nu, \mu' \nu'}} = \ldots = \frac{d A_{\mu \nu}}{f_{\mu \nu}}  = \frac{d t_{\alpha}}{\delta_{\alpha 1}}
\label{cheq2}
\end{eqnarray}
where
\begin{eqnarray}
    f_{\mu \nu} &=& \gamma A_{\mu \nu} - \sum_{\mu' \nu'} g_{\mu' \nu'} C_{\mu \nu, \mu' \nu'} \nonumber\\
    f_{\mu \nu, \mu' \nu'} &=& \gamma C_{\mu \nu, \mu' \nu'} - \frac{1}{2} \sum_{\mu'' \nu''} g_{\mu'' \nu''} \, C_{\mu \nu, \mu'' \nu''} \, C_{\mu' \nu', \mu'' \nu''},
    \label{fmunu}
\end{eqnarray}
where, $C_{\mu \nu, \mu' \nu'} = (1 + \delta_{\mu \mu'} \delta_{\nu \nu'}) B_{\mu \nu, \mu' \nu'}$, and $A_{\mu \nu}$ and $B_{\mu \nu, \mu' \nu'}$ are defined in eqs. \eqref{jpdfxxz1}.
A solution of the above equation for $t_1$ can be given as 
\begin{eqnarray}
t_1 = \sum_{\mu \nu, \mu' \nu'} q_{\mu \nu, \mu' \nu'} \int \frac{d B_{\mu \nu, \mu' \nu'}}{ f_{\mu \nu, \mu' \nu'}} +  \sum_{\mu \nu} q_{\mu \nu} \int \frac{d A_{\mu \nu}}{f_{\mu \nu}} + constant
\label{cheq3}
\end{eqnarray}
with $q_{\mu \nu, \mu' \nu'}$ and $q_{\mu \nu}$ as arbitrary constants dependent on initial conditions, and, as before, can be chosen to be $1$ for simplicity. 

With only $t_1$ varying and $t_2, \ldots, t_M$ remaining constant during the evolution, hereafter we will refer $t_1$ as $Y$ and suppress mention of $t_2, \ldots, t_M$.

\section{Schmidt eigenvalues Dynamics with changing system conditions} \label{schmidtEvol}

An eigenstate $U_{\nu}$ of the Hamiltonian $H$ corresponding to an energy $e_{\nu}$ can in principle be obtained by solving the eigenvalue equation $H \, U_{\nu} = e_{\nu} \, U_{\nu}$, $\nu=1 \to N$. Contrary to the eigenvalues, the behavior of an eigenfunction  depends on the basis in which the latter is represented. As the suitable basis for the bipartite entanglement analysis of an eigenfunction is the bipartite basis, we express $U_{\nu}$ in the $N=N_A N_B$-dimensional product basis  $|a_k b_l \rangle$ with $k=1 \to N_A, l=1 \to N_B$ with $C_{\nu; kl}$ as the  components: 
\begin{eqnarray}
U_{\nu} =\sum_{k=1}^{N_A}  \sum_{l=1}^{N_B} \; C_{\nu; kl} \; |a_k \, b_l \rangle. 
\label{uu}
\end{eqnarray}
With the components of $U_{\nu}$  now labelled by two indices, it is appropriate to represent them in a $N_A \times N_B$ matrix form $C_{\nu}$ referred as a state matrix. (Thus, in contrast to the eigenfunction matrix $U$ which consists of $U_{\nu}$, $\nu=1 \to N$ as its columns, $C_{\nu}$ represents a single eigenfunction $U_{\nu}$). With further analysis confined to a single eigenfunction, hereafter the subscript $\nu$ in  $C_{\nu}$ will be suppressed for clarity purposes unless necessary. 

\subsection{Dynamics of the eigenfunction components} \label{cp-formulation}

As the eigenvalue equation implies,  a randomization of $H$ results in fluctuations of the eigenvalues and eigenfunctions. The  joint probability density function (JPDF)  of the latter is related to the probability density of $H$ matrix as follows
\begin{eqnarray}
 P(e_1, \ldots, e_N, 
U_1, \ldots, U_N) \; \prod_{\nu=1}^N \left({\rm d}e_{\nu} {\rm d}U_{\nu} \right) = \rho(H) {\rm D}H,
\label{efh}
\end{eqnarray}
A complexity parameter based formulation of $\rho(H)$ along with the above relation can then be used to derive a similar formulation for the statistics of the eigenfunctions and eigenvalues. With our interest in a single eigenfunction dynamics in a bipartite basis, we confine the discussion here to the  joint probability density function (JPDF) of  the components $C_{kl}$ of $U_{\nu}$ and proceed as follows. The JPDF $P_c(C) \equiv P_c(C_{11}, C_{12}, \ldots, C_{NN})$ of the components $C_{kl}$ of $U_{\nu}$  is defined as
\begin{eqnarray}
P_{c}(C)  = \int \prod_{k,l=1}^{N_A, N_B} \delta(C_{kl}-C_{kl}(H)) \delta(C_{kl}^*-C_{kl}^*(H)) \; \rho(H) \; {\rm d}H.
\label{pu0}
\end{eqnarray}

Differentiating the above equation with respect to $Y$, followed by substitution of eq.(\ref{rhot}) and repeated partial integration then leads to 
\begin{eqnarray}
{\partial P_c  \over\partial \Lambda_{\psi}} =  \left(L +L^* \right) P_c
\label{pu} 
\end{eqnarray}
where 
\begin{eqnarray}
L \equiv \sum_{k,l} { \partial  \over \partial C_{kl}} \left(\sum_{m,n}  {\partial (\delta_{mk} \delta_{nl} - C_{kl} C_{mn}^*) \over \partial C^*_{mn}} + (N-1) C_{kl} \right)
\label{LL} 
\end{eqnarray}
  and $\Lambda_{\psi}$ is the rescaled evolution parameter
\begin{eqnarray}
\Lambda_{\psi}(e) = \chi \; (Y-Y_0) \; R_{local}^2
\label{lam0}
\end{eqnarray}
with $R_{local}(e)$ as a system-specific  energy scale around energy $e$ within which the eigenfunctions are correlated. The pre-factor $\chi$ is in general a  function of both $Y-Y_0$ and $e$; its exact form is not known so far and the only option left to us at this stage is to conjecture it based on intuition and verify its numerically as discussed later in section \ref{numerical} (also see Section {S}III of the Supplementary Material). We recall here that $Y-Y_0$ does not depend on the energy $e$ and the latter enters in $\Lambda_{\psi}(e)$ formulation only through $R_{local}$.

It is important to emphasize here the difference between  the mean level density of state $R_1(e)=\langle \sum_n \delta(e-e_n) \rangle$ and $R_{local}(e)$: in contrast to the former, $R_{local}$  corresponds to the weighted density of states, with weight corresponding to those eigenstates at energy $e$ which also occupy same basis space. We recall here that the eigenfunctions with neighbouring energies on the energy axis need not be spanning same part of the Hilbert space and therefore need not be correlated \cite{sup}. In earlier works on spectral statistics \cite{psand,psflat,psbe,pschiral,psbe2,tmchiral} and entanglement statistics in one-body systems \cite{shekhar2025single}, the measure $R_{local}$ was defined as  $R_{local} = {\xi^d \over N \Delta_e(e)}$ with $\Delta_e(e) = 1/R_1(e)$ as the ensemble averaged mean level spacing at the energy $e$ and $\xi^d(e)$ as the ensemble averaged localization length. While the rigorous route to determine the latter requires a detailed transfer matrix analysis, the technical complexity is often reduced by its approximation as $\xi ^d \approx \bk{I_2}^{-1}$ where $\bk{I_2(e)}$ is the ensemble averaged inverse participation ratio (IPR) at energy $e$. In previous studies \cite{psand, psflat, tmchiral} of single particle spectrum, the numerical analysis of the bulk spectral statistics based on the approximation was found to be consistent with $\Lambda_e$ based theoretical prediction. The relation $\xi ^d(e) \approx \bk{I_2(e)}^{-1}$ is however believed not to be well-applicable near the edge of the spectrum. In addition, with both single particle localization length as well as many-particle localization lengths playing important role in the wave-dynamics for many body systems, the previous definition of $R_{local}$ need not  be valid anymore. Due to lack of clear theoretical insights in this context, we follow the standard practice of numerical estimation. We find that in the present case, the average localization length ($\xi^d$) in the definition of $R_{local}$ is replaced by $N  \Omega_e$, i.e.,  $R_{local} = {\Omega_e \over  \Delta_e(e)}$  where $\Omega_e$ to a measure of correlation of the eigenstate at energy $E$ with the other eigenstates, the latter defined as 
\begin{equation}
 \Omega_e = \sum_{n=1}^{N_f} \sum_{k=1}^N \bk{|\psi_k| \, |\phi_{kn}| }
    \label{covar}
\end{equation}
where, $N_f$ is the number of eigenstates $\phi_n$, $n=1 \to N_f$, in the neighborhood $\delta e$ of the targeted eigenstate $\psi$ at energy $e$, used for the ensemble averaging.

As discussed  in \cite{psco, psand} in detail,  the eigenvalue dynamics  of the ensemble $\rho(H)$ with changing ensemble parameters can also be described by a mathematical formulation governed by a single function of all ensemble parameters. The latter, referred as the {\it {spectral complexity parameter}}, is defined as  $\Lambda_e(e) =  (Y - Y_0) \; R_{local}^2$ \cite{pseig, psco}. Here again the competition between $N$-dependence of  $Y-Y_0$ and $R_{local}$ leads to a critical spectral statistics at energy $e$ if 
\begin{eqnarray}
\Lambda_e^* \equiv \lim_{N \to \infty} \Lambda_e(e)= {\rm finite}.
\label{lamc}
\end{eqnarray}
A comparison of the above equation with eq.(\ref{lam0}) gives 
\begin{eqnarray}
 \Lambda_{\psi} (Y-Y_0; e) = \chi \; \Lambda_e(Y-Y_0; e),  
   \label{Lambda}
\end{eqnarray}
To distinguish it from $\Lambda_e$ and $Y-Y_0$,  hereafter we refer $\Lambda_{\psi}$ as the {\it {strength complexity parameter}}. The above relation implies significant correlations among eigenvalues and eigenfunctions in the critical regime and is in agreement with previous statistical studies of the complex Hamiltonians.

Eq.(\ref{pu})  describes  the $\Lambda_{\psi}$  governed evolution of the JPDF  of the state matrix elements $C_{kl}$  for a fixed system size $N$ starting from  for an arbitrary initial condition. Previously an evolution equation  for the state matrix elements representing an arbitrary {\it engineered quantum state} in a bipartite basis  was derived in \cite{psds1, psds2, psds3}. The distribution $P_c(C)$ in these studies was assumed to be a multiparametric Gaussian ensemble with independent matrix elements; assuming that the  state components are described only by  the first two moments and are uncorrelated; such a distribution for a state matrix ensemble follows  directly by invoking maximum entropy hypothesis. Interestingly the equation in the {\it engineered state case}  can again be written in the same form as  eq.(\ref{pu}) but the generator $L$ differs in important details: $L \equiv \sum_{kl} { \partial  \over \partial C_{kl}} \left( {\partial\over \partial C_{kl}^*} + C_{kl} \right)$. We note the missing correlation terms between components can not simply be introduced by consideration of a correlated Gaussian ensemble of state matrices.

\subsection{Dynamics of the Schmidt eigenvalues}

As mentioned in section \ref{state}, the entanglement entropy $R_1$ corresponds to the von Neumann entropy of the reduced density matrix $\rho_A = C C^{\dagger}$ and can be determined, in principle from the Schmidt eigenvalues.  Following from the above definition, a  randomization of $C$-matrix is expected to cause the fluctuations of $\lambda_n$, thereby making it relevant to derive a theoretical formulation of the JPDF $P({\lambda}) \equiv P(\{\lambda_i\}) \equiv P(\lambda_1, \ldots, \lambda_{N_A})$. A prior knowledge of the latter can then be used  to derive the ensemble average of $R_1$ as well as its higher order moments.

We proceed as follows. A comparison of eq.(\ref{pu}) with standard Fokker-Planck equation gives the $\Lambda$-dependent moments of the $C$-elements, i.e., $\langle C_{kl} \rangle, \langle C_{kl} C_{mn}^*\rangle$ and thereby the moments of the matrix elements $\rho_{A;mn}= \sum_{k=1}^{N_A} C_{mk} C_{nk}^*$ (details given in \cite{sup}). The latter along with   the second order perturbation theory for Hermitian matrix eigenvalues then leads to the  $\Lambda$-dependent moments of $\lambda_n$. Assuming Markovian dynamics for the latter, a substitution of their moments in standard Fokker-Planck equation then leads to  evolution equation for $P({\lambda})$ with $\Lambda_{\psi}$ as the evolution parameter,
\begin{eqnarray}
 \frac{\partial P}{\partial \Lambda_{\psi}}= {\mathcal L} P - {\mathcal L}_{cor} P
 \label{pdl0}
\end{eqnarray}
where
\begin{eqnarray}
{\mathcal L} &\equiv & \sum_{n=1}^{N_A} \left[\frac{\partial^2 (\lambda_n \; P)}{\partial \lambda_n^2} - \beta \frac{\partial}{\partial \lambda_n}\left( \sum_{m=1 \atop m \not=n}^{N_A} \frac{\lambda_n}{\lambda_n- \lambda_m} +  \nu_e - \eta  \lambda_n  \right)\right] \\
{\mathcal L}_{cor} &\equiv & \sum_{m, n = 1}^{N_A} \frac{\partial ^2}{\partial \lambda_n \lambda_m} \left[(\lambda_n \lambda_m)P\right]      
\label{pdl2}
\end{eqnarray}
with $\eta=\frac{N_A \, N_B}{2}$ and $\nu_e=\frac{N_A-N_B-1}{2}$. We note that, with presence of an additional term, i.e.,  ${\mathcal L}_{cor} P$, eq.(\ref{pdl0}) differs from the one derived in previous work for the engineered state ensemble \cite{psds1}. The additional term indicates a crucial difference from the previous study: it reflects  the effect of correlations among eigenfunction components present in a quantum state derived from a physical Hamiltonian; this aspect was not considered  in the engineered quantum states discussed in previous works \cite{psds1,psds2,psds3}.

\subsection{Critical Statistics} \label{crit}

The reduction of the multiparametric dynamics  to a single parametric one discussed in previous section is useful not only for technical purposes, but it also reveals a hidden web of connection as well as  infinite range of universality classes of the eigenfunction statistics among the eigenstates of those Hamiltonians which can be represented by $\rho(H)$. This can be further elucidated as follows.

For a fixed $N_A, N_B$, the solution of  Eq.(\ref{pdl0})  depends on  $\Lambda_{\psi}$ only, the latter  a function of ensemble parameters which in turn are governed by underlying system conditions.   This implies an analogy of the solutions  for two different states  if (i) both start  from statistically similar initial conditions, (ii) share same $\Lambda_{\psi}$ values, (iii) belong  to same or different  Hamiltonians represented by the ensemble $\rho(H)$, although their ensemble parameters can in general be different. In addition, as a change in the system parameters  for  finite $N$  can change $\Lambda_{\psi}$ continuously between $0$ and $\infty$, this predicts the existence of an infinite range of universality classes of the JPDF characterized by continuous values of $\Lambda_{\psi}$ between $0$ and $\infty$.

Additional important insight can be gained  by noting  that  $\Lambda_{\psi}$ itself is a function of $N$ through both $Y-Y_0$ and $\Omega_e$. A subtle competition between the $N$-dependence of the two scales, especially in large $N$ limit, can then play an important role in determining the statistics. For example, assuming $Y-Y_0 \propto N^{s}$ and $\Omega_e \propto N^q$, with $s, q$ as the system-dependent constants, gives $\Lambda_{\psi} \propto N^{s-2 q}$. As $ N \to \infty$, therefore, $\Lambda_{\psi} \to \infty$  for the system conditions leading to $s > 2 q$    and thereby to the uniformly distributed components of  the Haar-unitary state. Similarly, for the system conditions leading to $s < 2 q$, $\Lambda_{\psi} \to 0$ and the state ensemble remains stuck at its initial state. If however a specific combination of the prevailing system conditions lead to $s=2 q$, this renders $\Lambda_{\psi}$ independent of $N$; referring the corresponding $\Lambda_{\psi}$ value as $\Lambda_{\psi}^*$, it would then remain finite even in the $N \to \infty$ limit \cite{pseig}: 
\begin{eqnarray}
\Lambda_{\psi} ^* =\lim_{N \to \infty} \Lambda_{\psi} \not=0, \infty.
\label{almc}
\end{eqnarray}
The statistics at $\Lambda_{\psi} ^*$ remains therefore different from the two end points (i.e., $\Lambda_{\psi}=0$ and $\infty$) even in infinite size limit and can be referred as critical. Indeed,  based on the complexity of the system, more than one set of system parameters may exist, resulting in more than one $\Lambda_{\psi} ^*$ and thereby multiple  critical statistics intermediate between the initial state and Haar-unitary ensembles. In infinite size limit,  Eq.(\ref{pdl0}) therefore indicates  the existence of discrete universality classes for the eigenfunction statistics, each characterized by a distinct $\Lambda_{\psi} ^*$. The explicit appearance of size $N$  in Eq.(\ref{pu}) (in addition to its implicit appearance through $\Lambda_{\psi} $), 
however, suggests that the  statistics at $\Lambda_{\psi} ^*$ for  finite $N$ is different from that of infinite $N$.

Eq.(\ref{Lambda}) implies the existence of a finite size scaling and a multifractal behavior of the eigenfunction statistics at the critical point, with scaling exponents (referred to as critical exponents or multifractal dimensions) dependent on the system parameters \cite{pseig}. As $\Lambda_e$  is sensitive to system specifics, the critical (multifractal) exponents can vary from system to system.  The existence of a critical spectral statistics and multifractal behavior of the eigenstates however requires a specific set of system conditions conspiring with each other and leading to a size-independent $\Lambda_{\psi}$ (a more detailed discussion of $\Lambda_{\psi}$ is included in the supplementary material \cite{sup}).

As eq.(\ref{Lambda}) indicates, the rescaling of $Y-Y_0$ leads to $e$-dependence of $\Lambda_{\psi}$, thereby implying a lack of translational invariance in the statistical behavior along the spectral axis (also referred as the non-stationarity).

\section{Complexity parameter formulation of the entanglement measures} \label{evolmeas}

In this section, we derive the complexity parameter formulation of the average behavior as well as the variance  of the standard  measure used to quantify entanglement, namely the von Neumann entropy. To avoid cluttering of  the symbols, we hereafter suppress the subscript $\psi$ from  $\Lambda_{\psi}$  and restore it only if necessary for clarity.

\subsection{von Neumann entropy statistics}

The ensemble averaged von Neumann entropy $\langle R_1 \rangle$ can be defined as
\begin{eqnarray}
\langle R_1 \rangle= \int  \left(- \sum_n \lambda_n \log \lambda_n \right) \; P(\lambda) \; {\rm D} \lambda.
\label{rvn1}
\end{eqnarray}

Differentiating the above equation with respect to $\Lambda$, followed by substitution of eq.(\ref{pdl0}) and simplifying by partial integration (details discussed in \cite{psds1} and also in \cite{sup}), we have

\begin{equation}
\pdy{\bk{R_1}} = \alpha + \frac{N_B}{2} \bk{R_0} - \frac{N_A N_B}{2} \bk{R_1}. \label{evoleqr1}
\end{equation}
with $\alpha= 1 - \frac{N_A(N_A+1)}{2} $ and $R_0 = - \sum_n \log \lambda_n$ with $\bk{R_0}$ as its ensemble average.

The above equation describes the $\Lambda$-governed growth of average von Neumann entropy, from an arbitrary initial state at $\Lambda=0$ (equivalently $Y=Y_0$). As $R_0 \to \infty$ under separability condition (i.e. $\lambda_n =0$ for all $n=1 \to N$ except one of them), the above implies a rapid variation of  $\pdy{\bk{R_1}}$ for $\Lambda$ near $0$ if the initial state ensemble  is separable. Further, with $\pdy{\bk{R_1}} \to 0$ and $\bk{R_0} \to N_A \log N_A + {N_A^2 \over 2 N_B}$  as $\Lambda \to \infty$ \cite{psds1}, the above equation gives  $\bk{R_1} = \log N_A -{N_A \over 2 N_B}$  for large $\Lambda$. The latter agrees with the Page limit \cite{page1993average} for the ergodic states; (we recall that the state $\Psi$ approaches ergodic limit as $\Lambda \to \infty$). A change from $\pdy{\bk{R_1}} \to 0$ to $\infty$ is expected to take place at a finite $\Lambda$ value when $\bk{R_0}$ becomes finite and is of the same order as that of $\bk{R_1}$. Indeed, as our numerics indicated, the change occurs where the correlation between $R_0$ and $R_1$ is maximum.

A general solution for eq.(\ref{evoleqr1}) for arbitrary $\Lambda$ and an arbitrary initial condition can be given as 
\begin{equation}
    \bk{R_1}(\Lambda) = \bk{R_1}(0)  \; e^{-\frac{N\Lambda}{2}} + \frac{2 \alpha}{N} \left(1- e^{-\frac{N\Lambda}{2}}\right) 
    +\frac{N_B}{2} \, e^{-\frac{N\Lambda}{2}} \, \int_0^{\Lambda}  \bk{R_0} \,  e^{\frac{N\Lambda}{2}} \, {\rm d}\Lambda
 \label{avgr0}
\end{equation}
The presence of $\bk{R_0} = \bk{R_0}(\Lambda)$ in the above solution however rules out  the choice of  separability limit as an initial condition to determine $ \bk{R_1}(\Lambda) $. This is because, with one of the eigenvalues as $1$ and rest zero in the above limit,  $R_0=-\sum_n \log \lambda_n \to \infty$. It is however possible to choose a weak separability limit as an initial condition at $\Lambda=0$; this can be explained as follows.   From its  definition, $R_0$  become singular even if only one of the eigenvalues is zero.  Starting from the separability limit, a change of system conditions may however lead to a change of eigenvalues, with the unit eigenvalue now decreasing and rest of them increasingly becoming nonzero. Indeed, $R_0$ become finite as soon as one of the eigenvalues is $1-N_A^{-q}$ and rest of them are $\sim N_A^{-(q+1)}$ with $q > 1$. This gives $R_0 \sim (q+1)  N_A \log N_A$ and $R_1 \sim N_A^{-q}(1-N_A^{-q}) + q N_A^{-q} (N_A-1) \log N_A$. Thus, taking the initial condition at $\Lambda=0$ as one of the eigenvalues $1-N_A^{-q}$ and rest of them as $\sim N_A^{-(q+1)}$,  we have $\bk{R_1}(0)=0$ with $q > 1$. 

As clear from the above, $\bk{R_1}(\Lambda)$  rapidly changes from  $0 \to \log N_A$ near $\Lambda \approx 2/N$. Indeed, this is expected, with the $\Lambda$-value corresponding to a change of $R_0$ from $\infty$ to a finite value  and $R_1$ from $0$ to its maximum limit,   and can be explained as follows.    As mentioned above,  only one of the eigenvalues, say $\lambda_1$, is one and rest are zero at the separability limit. With changing system conditions however the zero eigenvalues evolve increasingly becoming nonzero; second order perturbation theory of Hermitian matrix spectrum then gives  $\langle \delta \lambda_n \rangle = \beta \left[- (N-1) \,  \lambda_n  + \sum_{m (\neq n)} \frac{\lambda_n + \lambda_m}{\lambda_n - \lambda_m} \right] \; \delta \Lambda$ (discussed in section III of supplemental material). For a zero eigenvalue at $\Lambda=0$, this implies $\langle \delta \lambda_n \rangle \sim {\delta \Lambda\over \Delta_{\lambda}}$ or equivalently $\lambda_k \sim  {\Lambda\over \Delta_{\lambda}} = {2\over N \Delta_{\lambda}}$ with  $k=2 \ldots, N_A$ with $\Delta_{\lambda}$ as the mean level spacing of the Schmidt eigenvalues: $\Delta_{\lambda} \sim 1/N$; (we recall that the measure $\Delta_e(e)$ introduced above eq.(\ref{covar}) referred to Hamiltonian spectrum with $e$ as the energy).  For $\Lambda \ll \Delta_{\lambda}$, this gives, from the  definitions, $R_0 \sim - N_A \log \left({\Lambda\over \Delta_{\lambda}} \right)$ and $R_1 \sim \left(1-{\Lambda\over \Delta_{\lambda}}\right) {\Lambda\over \Delta_{\lambda}}  - {N_A \Lambda\over \Delta_{\lambda}} \log \left({\Lambda\over \Delta_{\lambda}} \right)$, thus implying $R_0$ remaining very large and $R_1$ negligible.  A change however occurs $\Lambda \sim \Delta_{\lambda}$ with $R_0$ becoming finite and $R_1$ large for $\Lambda \gg \Delta_{\lambda}$. This is also confirmed by our numerical analysis discussed in section \ref{numerical}.

\begin{figure}
    \centering
    \begin{subfigure}[t]{0.49\textwidth}
        \centering
        \includegraphics[width=\textwidth]{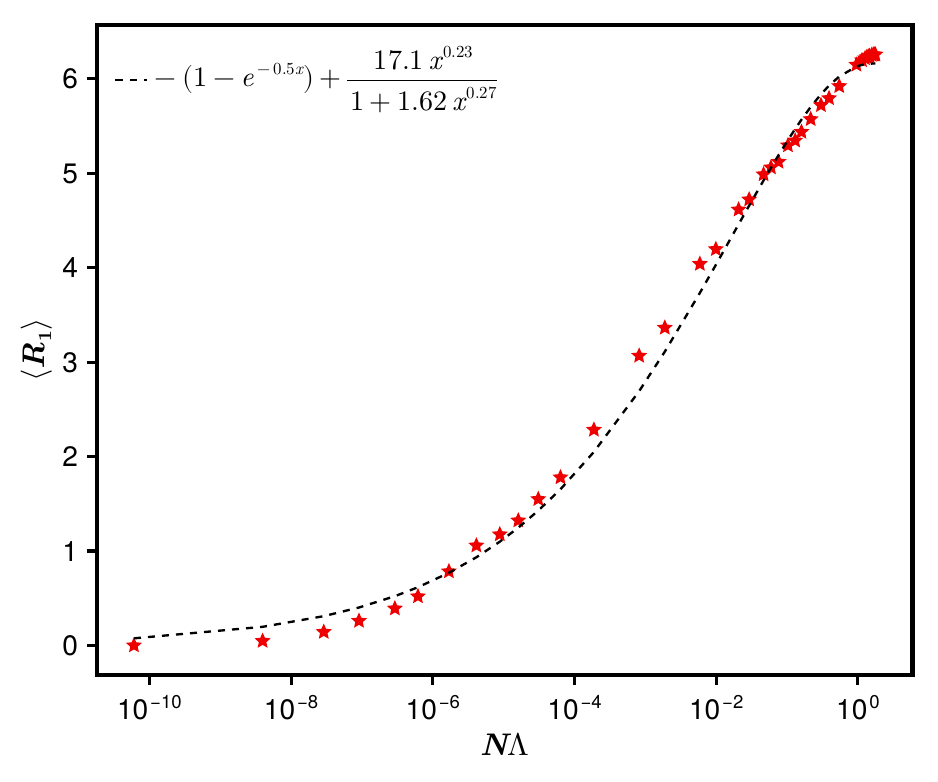}
    \end{subfigure}
    \hfill
    \begin{subfigure}[t]{0.49\textwidth}
        \centering
        \includegraphics[width=\textwidth]{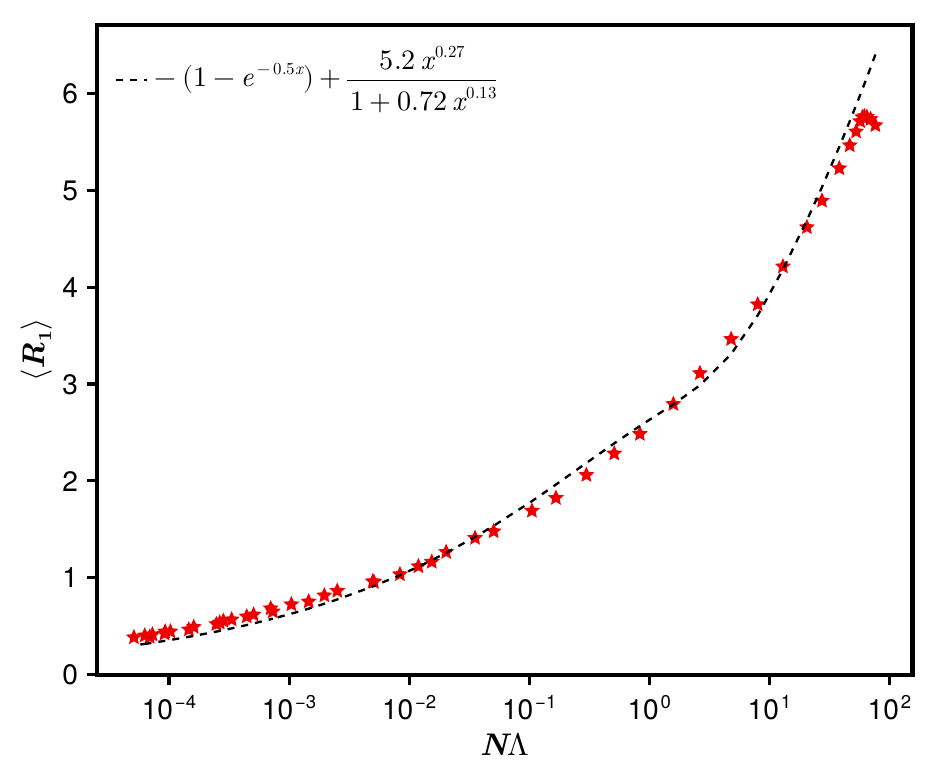}
    \end{subfigure}
    \caption{\textbf{Fit of Average von Neumann entropy.} The figure shows the fitting of numerical result of $\bk{R_1}$ for (left) QREM and (right) RFHM. The corresponding fit functions are also displayed in the legend with $x \equiv N \Lambda$. Interestingly, we find that the power in QREM is changing at $x=1/N$ (indicated by fitting of two different powers but same function), however, this behavior is not apparent in the RFHM case.}
    \label{fig:fit}
\end{figure}

To evaluate the integral in eq.(\ref{avgr0}), we also need  a prior knowledge of $\Lambda$-dependence of $\bk{R_0}$ for finite $\Lambda$. Indeed, following the similar steps as  for   $\bk{R_1}$, a $\Lambda$-governed evolution equation for  $\bk{R_0}$ can also be derived 
\begin{equation}
    \pdy{\bk{R_0}} = \bigg(\beta \nu (N_A+1)  +1 \bigg) \bk{R_{-1}} +  \bigg(\frac{\beta N}{2} -1 \bigg) N_A 
     \label{diffr0}
\end{equation}
where $R_{-\alpha} = \sum_n {1\over \lambda^{\alpha}} = (-1)^{\alpha}  \sum_{n}  \bigg({\partial R_0 \over \partial \lambda_n}\bigg)^{\alpha}$. But the above depends on a negative moment of the Schmidt eigenvalues  and any attempt to solve it further leads to a set of hierarchical equations for negative moments with no currently available solution. To overcome the  technical difficulty, the only option available at this stage is to apply numerically derived insights to conjecture $\Lambda$-dependence of  ${\bk{R_0}}$. We conjecture  ${\bk{R_0}}\approx {a \; N \; \Lambda^b \over 1 + c \Lambda^d} $ with $b, d$-values dependent on $\Lambda$; $b \sim d$ for correct limiting behaviour of ${\bk{R_0}}$ in $\Lambda \to \infty$ limit. The latter's substitution in the integral of eq.(\ref{avgr0}) gives for the leading order term, in limit $N \to \infty$ but for small $\Lambda$, $   \bk{R_1}(\Lambda) = \frac{2 \alpha}{N} \left(1- e^{-\frac{N\Lambda}{2}}\right) + {a \; N \; \Lambda^b \over 1 + c \Lambda^d}$. A comparison  of the numerical obtained small $\Lambda$ behavior for $\bk{R_1}$ for both QREM (at $E=0$) and RFHM (at $D=1$) with the fitted function is displayed in fig. \ref{fig:fit}. Although a dependence on three unknown parameters $b, c, d$ makes the above conjecture a weak one, nonetheless the validity of the conjectured form for the eigenstates of two very different Hamiltonians lends it some credence. 

\begin{figure}[h]
    \centering

   \includegraphics{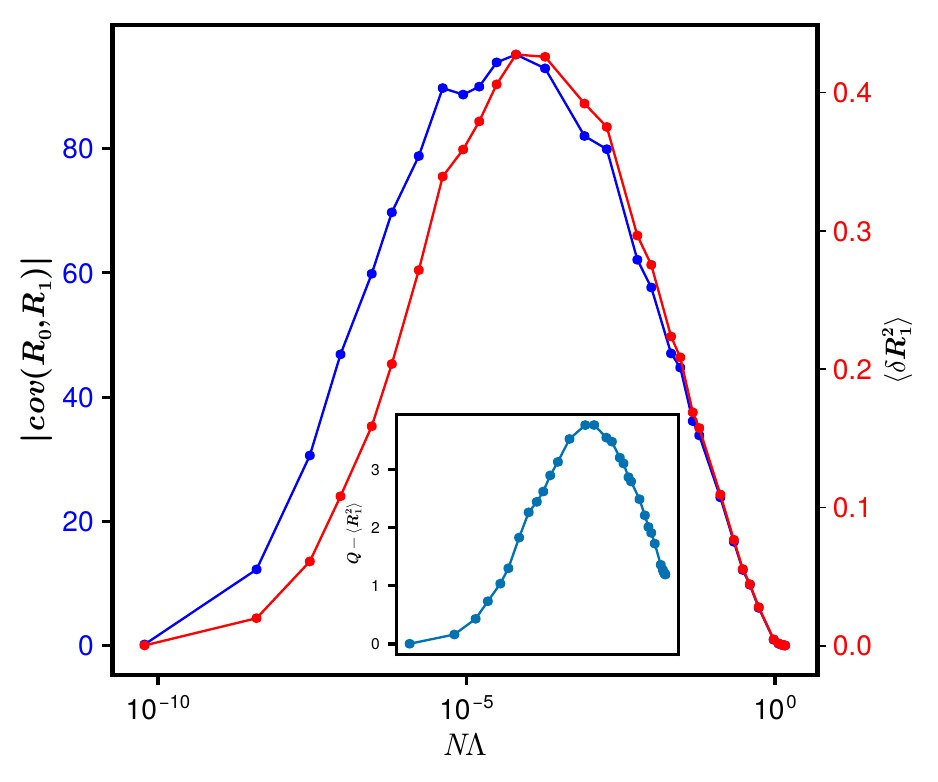}
   \caption{\textbf{Analysis of the RHS of the eq. \eqref{evoleqr2}.} We compare the $|\text{cov}(R_0, R_1)|$ (blue) and the $\bk{\delta{R_1^2}}$ (red) for the eigenstates of the QREM at $E=0$ for $L=14$, and with the x-axis on a log scale. As can be seen, the quantities are qualitatively quite similar, and that in the ergodic regime the former tend to zero. The  diverging variance of $R_1$ is therefore a consequence of the diverging covariance $\text{cov}(R_0, R_1)$. (inset) We show the dynamics of $Q - \bk{R_1^2}$ (ref. eq. \eqref{evoleqr2}), which is also diverging but $\sim 1$ as $\Lambda \to \infty$, and hence dominates the covariance term in that limit. Similar results are obtained for the RFHM.}
    \label{fig_cov}
\end{figure}

Proceeding along the same route, we  derive the $\Lambda$ governed evolution of the variance of the von Neumann entropy, defined as $\langle \delta R_1^2\rangle = \langle R_1^2\rangle - \langle R_1 \rangle^2$ (the details discussed in \cite{sup}). 
\begin{equation}
\pdy{\bk{\delta R_1^2}} = 2(Q - \bk{R_1^2}) + N_B \, \text{cov}(R_0, R_1) - N\bk{\delta R_1^2} \label{evoleqr2}
\end{equation}
where  $Q \equiv \bigg \langle \sum_n \lambda_n (\log \lambda_n)^2 \bigg \rangle$ and $\text{cov}(R_0, R_1)$ refers to the covariance of $R_0$ and $R_1$:  $\text{cov}(R_0, R_1)\equiv \bk{R_0 R_1}-\bk{R_0}\bk{R_1}$.  A general solution of eq.(\ref{evoleqr2}) for large $N$,  arbitrary $\Lambda$ and arbitrary initial conditions can now be given as 
\begin{equation}
   \bk{\delta R_1^2}(\Lambda) \approx   \; e^{-N\Lambda} \, \bigg(\bk{\delta R_1^2}(0)  +   \int_0^{\Lambda} \bigg(2(Q - \bk{R_1^2}) + N_B \; \text{cov}(R_0, R_1) \bigg) \,  e^{N\Lambda} \, {\rm d}\Lambda \bigg)
 \label{avgr2}
\end{equation}
In the large $N$ limit and for small $\Lambda$,  the term $Q-\bk{R_1^2} \ll \text{cov}(R_0, R_1)$ and the evolution of $\bk{\delta R_1^2}$  is dominated by the term  $\text{cov}(R_0, R_1)$.  Consequently, the above solution  requires a prior knowledge of $\Lambda$-dependence of $\text{cov}(R_0, R_1)$. For initial state at $\Lambda=0$ chosen as separability limit, $\bk{\delta R_1^2}(0)=0$. A small change in $\Lambda$ from $0$ to, say $\Lambda_1 \sim O(1/N)$ causes  a rapid increases and thereafter decay of  $|\text{cov}(R_0, R_1)|$; eq.(\ref{avgr2}) then gives 
\begin{equation}
   \bk{\delta R_1^2}(\Lambda) \approx   c_0 \; N_B \; e^{-N(\Lambda-\Lambda_1)}
 \label{avgr3}
\end{equation}
where $c_0$ is the value of $|\text{cov}(R_0, R_1)|$ at $\Lambda=\Lambda_1$. The above prediction is also confirmed by our numerics displayed in fig. \ref{fig_cov}. Indeed, we find the behavior of $|\text{cov}(R_0, R_1)|$ qualitatively the same as $\bk{\delta R_1^2}$ (fig. \ref{fig_cov}), including their divergence.

In the large $\Lambda$ limit,  with $R_0 \sim N_A \, R_1$,  the $\text{cov}(R_0, R_1)$ almost vanishes and  the term  $Q-\bk{R_1^2}$ dominates, and eq.(\ref{avgr2}) now gives 
\begin{equation}
   \bk{\delta R_1^2}(\Lambda) \approx   2\frac{(Q - \bk{R_1^2})}{N} \, (1-e^{-N\Lambda}) =   2\frac{(Q - \bk{R_1^2})}{N} \; { \bk{ R_1}(2 \Lambda) \over  \bk{ R_1}(\infty)}
 \label{avgr3}
\end{equation}
In the ergodic limit $\Lambda \to \infty$, $Q - \bk{R_1^2} \approx 1$, then the above gives $\bk{\delta R_1^2} \sim \frac{2}{N}$ and is consistent with previous studies \cite{psds3,nadal2011statistical,bianchi2019typical}.

\subsection{R\'enyi entropy statistics}

Another way to quantify entanglement is through the second R\'enyi entropy $R_2 \equiv -\log \left(\sum_n \lambda_n^2 \right)$, with its ensemble averaged defined as
\begin{eqnarray}
\langle R_2 \rangle= \int  \left(- \sum_n \lambda_n^2 \right) \; P(\lambda) \; {\rm D} \lambda.
\label{re2}
\end{eqnarray}
Proceeding exactly as in the case of $R_1$ again leads to a complexity parameter formulation 
for $\langle R_2 \rangle$. As the intermediate steps are essentially similar and differ only in details, 
here we mention only the final forms of the evolution equations for the average and the variance 
of $R_2$ for large $N$ (see the supplementary material for details),

\begin{equation}
    \frac{\partial \bk{R_2}}{\partial \Lambda} = 4 \bk{\frac{S_3}{S_2^2}} - 2(N_A - \nu) \bk{\frac{1}{S_2}} - 8 \bk{\frac{S_4}{S_2^2}} + (N+6),
    \label{pdyr2a}
\end{equation}
where, $S_k = \sum_n \lambda_n^k$. For $N \gg 1$, the above can be approximated as 
\begin{equation}
    \frac{\partial \bk{R_2}}{\partial \Lambda} \approx N - 2 (N_A - \nu) \bk{\frac{1}{S_2}}.
    \label{pdyr2b}
\end{equation}
where, $S_2 = \sum_n \lambda_n^2$ is the purity. A general solution for the above equation for arbitrary $\Lambda$ and an arbitrary initial condition can be given as 
\begin{equation}
    \bk{R_2}(\Lambda) = N \Lambda - 2 (N_A - \nu) \;   \int_0^{\Lambda} \bigg\langle{\frac{1}{S_2}} \bigg\rangle  \; {\rm d}\Lambda
 \label{avgr6}
\end{equation}

We note that, in contrast to  $R_0$ appearing in $\bk{R_1}$-case,  the integral in eq.(\ref{avgr6}) is now well-defined in the separability limit with $S_2 =1$.  To evaluate the integral however  a  prior knowledge of $\bk{1/S_2}$ is again required and again leads to a  set of hierarchical equations involving sums of the negative moments of the Schmidt eigenvalues.

The $\Lambda$-governed evolution equation for the variance of $R_2$ can similarly be derived. We have, for large $N$,
\begin{equation}
    \frac{\partial \bk{R_2^2}}{\partial \Lambda} \approx 8 \bk{\frac{R_2 S_3}{S_2^2}} - 4 (N_A - \nu) \bk{\frac{R_2}{S_2}} - 16 \bk{\frac{R_2 S_4}{S_2^2}} + 2 N \bk{R_2}.
    \label{pdyr2c}
\end{equation}
Further using  $\frac{\partial \bk{\delta R_2^2}}{\partial \Lambda} = \frac{\partial \bk{R_2^2}}{\partial \Lambda} - 2 \bk{R_2} \frac{\partial \bk{R_2}}{\partial \Lambda}$, we have from eqs. \eqref{pdyr2a} and \eqref{pdyr2c},
\begin{equation}
    \frac{\partial \bk{\delta R_2^2}}{\partial \Lambda} = 8 \, \text{cov}(R_2, \frac{S_3}{S_2^2}) - 16 \, \text{cov} (R_2, \frac{S_4}{S_2^2}) - 4 (N_A - \nu) \, \text{cov}(R_2, \frac{1}{S_2}).
\label{avgr7}    
\end{equation}
As the above equation indicates, the evolution of the variance of $R_2$ is dominated by the covariance $\text{cov}(R_2, \frac{1}{S_2})$. Thus, $S_2^{-1}$ plays the same role for $R_2$, as $R_0$ in the case of $R_1$.

While an exact functional dependence of $\bk{R_n}$ and $\bk{\delta R_n^2}$ for $n=1,2$ on $\Lambda$ are not available so far,  Eqs.(\ref{avgr0}, \ref{avgr2}, \ref{avgr6}, \ref{avgr7}) clearly indicates  an evolution  of $\bk{R_n}$ and $\bk{\delta R_n^2}$
governed by a single ensemble parameter that contains information about all system parameters. The above prediction is consistent with our numerical analysis of two many body Hamiltonians, namely, the QREM and the RFHM, discussed in the next section.

\section{Numerical Analysis} \label{numerical}

A typical quantum state of a many-body Hamiltonian $H$ depends in general  on many system parameters governing its various physical attributes. For an ensemble density $\rho(H)$ to be an appropriate representation of the statistical behaviour of $H$,  the ensemble parameters must depend on the system parameters (as elucidated by the  examples  in section \ref{density}). Thus, a variation of any of the system parameters is expected in general to cause a variation of the ensemble parameters and thereby the entanglement statistics.  Based on our theoretical prediction, however, the evolution of the entanglement statistics is governed only by two parameters, namely, $\Lambda$ and $N$ and not by the specific details of the ensembles parameters. Different states originating from similar initial conditions are then not only predicted to follow the similar paths in terms of $\Lambda$, they also correspond to same entanglement statistics if their $\Lambda$-values coincide. This indicates a potential classification of the quantum states of a given Hamiltonian (or different Hamiltonians subjected to same global symmetries and conservation laws) in non-ergodic universality classes characterized by the complexity parameter $\Lambda$. The ergodic universality class in this classification is characterized by $\Lambda \to \infty$. The deep significance of our theoretical claims makes it necessary to verify them numerically. For this purpose,  we consider the  Hamiltonian described by eq.(\ref{qrem})  taken from the ensemble in eq.(\ref{rhog}) and eq.(\ref{xxz}) taken from the ensemble in eq.(\ref{jpdfxxz0}). 
 
A physical Hamiltonian has in general many eigenstates, each characterized by  corresponding eigenvalue. Consequently, for a given ensemble of Hamiltonians, there are many state ensembles, each representing a specific eigenstate and characterized by  the state complexity parameter $\Lambda = \Lambda(e, Y)$ where $Y$ is the ensemble complexity parameter at energy $e$; (alternatively stated, each point of the Hamiltonian spectrum corresponds to a state ensemble). An important point worth re-emphasizing here is as follows. With $\Lambda$-dependent on energy range, the quantum states of the Hamiltonian $H$ for different energies correspond  in general to different $\Lambda$ values, although each belongs to the same fixed set of system parameters, thus leading to same $Y$. If however the system parameters are varied, the entanglement measures for each state evolve through an analogous path lying between separability limit to maximum entanglement (their rates of evolutions however may vary).

The determination of an average measure, e.g., $\bk{R_1}$ for an eigenstate say of energy $E$ requires, in principle, an averaging over the corresponding state ensemble only. The numerical analysis however requires an averaging over the neighbouring state ensembles too. This can be explained as follows. A state ensemble  $P_c(C)$  is obtained numerically by an exact diagonalization of an ensemble of Hamiltonians;  $P_c(C)$  then in principle corresponds to the set of  eigenstates of energy $E$ taken from each Hamiltonian of the  ensemble  $\rho(H)$. But as the spectrum locally  fluctuates from one Hamiltonian to the other, it is not possible in general to pick the eigenstates with exactly same energy $E$. The state ensemble is then numerically obtained by permitting a small fluctuation of energy, i.e., by considering the eigenstates within an energy range $e \pm \delta e$ from each Hamiltonian. Here  $\delta e$ is an optimized range, permitting consideration of only those states in $e \pm \delta e$ range which share same $\Lambda$. As the latter depends on $\Delta_e(e)$, the range $\delta e$ should be chosen smaller than  $\Delta_{e}(e)$ to ensure that $\Lambda$ remains almost same for all the sample states in the ensemble.

For numerical determination of the eigenstates, we use the standard \textit{shift-invert} diagonalization technique  \cite{pietracaprina2018shift}. An efficient implementation of the technique and efficient computation for large systems leveraging MPI techniques is ensured by  utilizing the SLEPc library in C \cite{slepc} for our numerical codes. To quantify entanglement of an eigenstate, we consider the system represented by $H$ divided into two halves, say $A$ and $B$ and calculate $H$ matrix in their product basis. The information is then used to determine the von Neumann entropy ($R_1 = -tr(\rho_A \log \rho_A)$) of the reduced state of one half of the system, i.e., $L_A = L/2$; (we use $\log$ base $2$ to numerically calculate the entropy). The average of the measures are determined by both spectral and ensemble averaging. The specific details for each case are discussed below.

\subsection{QREM} \label{QREM}

In case of QREM, the Hamiltonian $H$ has two system parameters, namely the transverse field strength $\Gamma$, which depends on $b$, and system size $L$ and both appear in the ensemble density eq.(\ref{rho1}) through ensemble parameters. With the initial condition $b = 0$, substitution of the latter in eq.(\ref{yparam}) leads to 
\begin{equation}
    Y-Y_0 = -\frac{1}{2(N + 1) \, \gamma} \, \sum_{r=0}^{L-1}\ln \bigg | 1 - \frac{2\gamma}{1 + (\frac{2^r}{b})^2} \bigg|.
    \label{yqrem}
\end{equation}
The $\sum_{r=0}^{L-1}$ above arises from the basis pairs $|k\rangle$ and $|l\rangle$ at a unit Hamming distance; the later correspond to the distance $|k-l| = 2^r$ in the basis space, with $r = 0 \cdots L-1$. Further, as $\gamma$ is arbitrary (related to the variance of the matrix elements in the ergodic limit), we choose $\gamma=1/2$ in QREM numerical calculations.

As mentioned in the section \ref{diff_density}, the condition $\frac{\partial \rho}{\partial t_{\alpha}} = 0 \quad \forall \; \alpha >1$ implies $t_{\alpha}$ as the constants of evolution, obtained by solving eq. \eqref{cheq1}. For example, for an arbitrary choice of differentials in the eq. \eqref{cheq1}, we have, $\frac{d v_{\mu \nu}}{f_{\mu \nu}} = \frac{d v_{k l}}{f_{kl}}$, with corresponding solution as $\log(\frac{g_{\mu \nu}- 2 \gamma v_{\mu \nu}}{g_{kl}- 2 \gamma v_{kl}}) = \text{const}.$; the constant so obtained can be chosen as one of the $t_{\alpha}$, for $\alpha > 1$. For example, for the QREM, it is clear from the eq. \eqref{qrem2}, that for a specific combination of pairs, such that Hamming distance between $\mu, \nu$ and $k, l$ is one, $v_{\mu \nu} = v_{kl}$. This gives $\log(\frac{1 - 2 \gamma v_{\mu \nu}}{1 - 2 \gamma v_{kl}}) = 0$; $t_{\alpha}$ can then be chosen to as $t_{\alpha} = \frac{1 - 2 \gamma v_{\mu \nu}}{1 - 2 \gamma v_{kl}} = 1$. As this can be done for several combinations of pairs, $t_{\alpha}$ can be chosen to be $1 \; \forall \alpha$. (see also \cite{psds2} for a more elaborate discussion.)

We exactly diagonalize the Hamiltonian in eq. \eqref{qrem} for several values of the free parameter $b$ ranging from 0, corresponding to a localized state, to a sufficiently large value such that the system reaches to the ergodic regime. But, as stated above, each $b$ value for a fixed $L$  leads to many state ensembles characterized by  $\Lambda(e)$; the energy $e$ can then be used as a free parameter too. This in turn gives us a state ensemble with two free parameters $b$ and $e$ for a fixed $L$, and we can now seek whether the ensemble averaged measures, e.g., $\langle R_1 \rangle$ indeed coincides quantitatively for different pairs of $b$ and $e$ but same $\Lambda$.

As mentioned in the beginning of this section, it is necessary to consider the behaviour of entanglement measures subject to both ensemble and spectral averages. For this purpose, we consider the average of $R_1$ and its variance  over various disorder realizations and over about $1\%$ of the total eigenstates per realization in the neighborhood of energy $e=E$, with corresponding $\Lambda$ determined from eq. \eqref{Lambda}.

To proceed further,  we need to determine $\Lambda$ from eq. (\ref{Lambda}). Besides $Y-Y_0$, this requires a prior knowledge of local mean level spacing $\Delta_e(e)$ as well as   $\Omega_e(e)$ as a function of free parameters. In absence of a theoretical formulation, we determine both $\Delta_e$ numerically, $\Omega_e$ is again determined using the relation eq. \eqref{covar}, and $\chi$, which is also determined numerically and is listed in the Table \ref{tab:chi} for different cases

\begin{table}[b]
\centering
\caption{\textbf{$\chi$ and $\Lambda_e$ for different models and measures:} We display the numerically verified form of $\chi$ and the $\Lambda_e$  for the QREM and the RFHM for the different statistical measures, viz., the average and the variance of the entanglement entropies. These forms are used to determine 
$\Lambda = \chi \; \Lambda_e(e)$  which appears in all the figures displayed in the paper. }
\begin{tabular}{lcccc}
\hline\hline
Model & Measure & $\chi$ & & $\Lambda_e(e)$ \\
\hline
\multirow{2}{*}{QREM} & $\bk{R_n}$ & $\frac{\Delta_e}{\bk{I_2}}$ & & \multirow{2}{*}{$-\frac{\Omega_e^2}{2 N \gamma \Delta_e^2} \sum_{r=0}^{L-1} \ln |1 - \frac{2\gamma}{1+(\frac{2^r}{b})^2}| $} \\
 & $\bk{\delta R_n^2}$ & $(\Delta_e)^{-0.4}$ & & \\
\hline
\multirow{2}{*}{RFHM} & $\bk{R_n}$ & \multirow{2}{*}{$\Delta_e^{3.1} \, \Omega_e^{1.5}$} & & \multirow{2}{*}{$\frac{\Omega_e^2}{\gamma \, \Delta_e^2}  \ln\left(\frac{h_0^4}{h^4} \frac{D}{D_0}\right) $} \\
 & $\bk{\delta R_n^2}$ & & & \\
\hline\hline
\end{tabular}
\label{tab:chi}
\end{table}

\begin{figure}
    \centering
    \includegraphics[width=\textwidth]{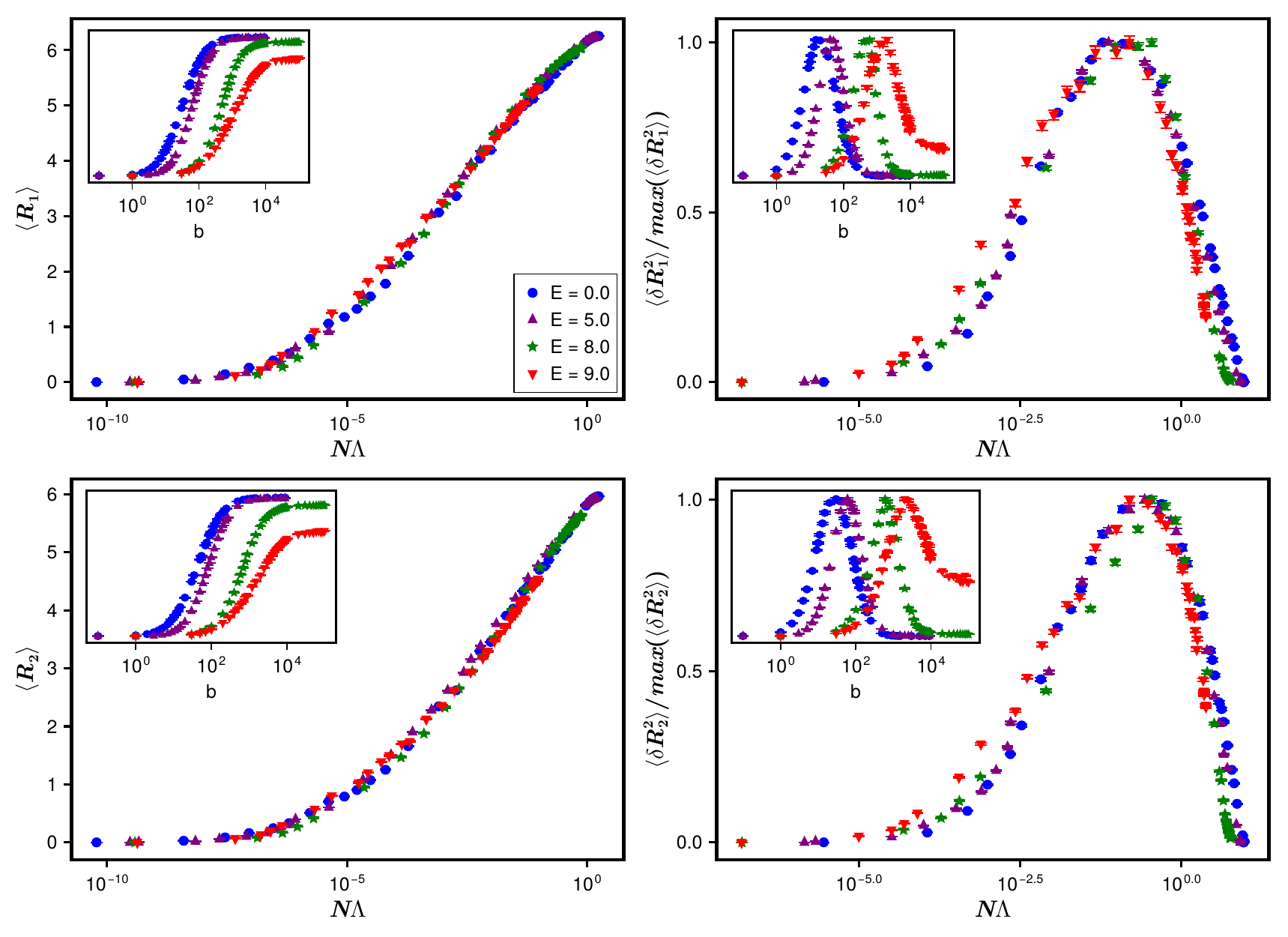}
    \caption{\textbf{Average and variance of von Neumann Entropy and the second R\'enyi entropy in the QREM.} The dynamics of the average and the variance of the (top) von Neumann entropy $R_1$, and (bottom) the second R\'enyi entropy ($R_2$) for the system size $L = 14$ over $1000$ disorder realizations, and $200$ eigenstates per realizations at different energy scales ($E$) is shown. The variance in each case has been rescaled by the maximum for a better comparison. As can be seen, the evolution curves for different energies overlap when plotted with $N \Lambda$, as opposed to with the system parameter $b$ (inset). The $\Lambda$ for the respective measures is shown in the Table \ref{tab:chi}.}
    \label{fig_r1_r2:qrem}
\end{figure}

Figs. \ref{fig_r1_r2:qrem} show the evolution of the average and the variance of the von Neumann entropy ($R_1$) and the second R\'enyi entropy ($R_2$) for a fixed system size, $L=14$, but for different energies $E$. We average over $1000$ disorder realizations and $200$ eigenvectors per realization ($N_f$) for various off-diagonal disorder parameter $b$ and energy levels $E$. As shown in the Table \ref{tab:chi}, $\chi$ for different statistical measures turns out to be different; this is consistent  with the theoretical prediction in \cite{pseig}.   Table \ref{tab:chi} also display the mathematical form of $\Lambda$ used in the figures. As is clear from the figures, while the evolution paths of the entanglement statistics, at different energy scales and with $b$ as the independent variable, are visibly distinguishable (see inset),  their difference vanishes with curves collapsing onto each other with $N \Lambda$ as the independent variable. We also note that,  the average $R_n$ for the states at different energies saturates to different  ergodic limits, with those near the spectral edge approaching a lower $R_n$ value. This behavior is consistent with $\Lambda$ formulation: at higher energy,  the mean level density decreases sharply, resulting in a smaller $\Lambda$ and hence lower entanglement.

\subsection{RFHM} \label{RFHM}

Our numerical analysis in this case is based on the Hamiltonian $H$ in eq.(\ref{xxz}) with  $J = 1$, $h_j$ as the Gaussian disorder with mean zero, and $\rho(H)$ given by eq.(\ref{jpdfxxz0}). A substitution of values given by eq.(\ref{jpdfxxz1}) in eq.(\ref{cheq3}) and further using eqs. \eqref{fmunu}, eq. \eqref{cheq3} can be approximated in the large $L$ limit to give (see \cite{sup} for details)

\begin{equation}
    Y-Y_0 \approx \frac{1}{\gamma} \log\left(\frac{h_0^4}{h^4}\frac{D}{D_0}\right),
    \label{yxxz}
\end{equation}
where, $h_0$ and $D_0$ correspond to initial conditions $h_0 \gg 1$ and $D_0 = 1.0$, and in the numerical calculations we take $\gamma=1$. As $H$ in this case conserves the total spin in the z-direction ($S_z^{total} = \sum_{i=1}^L S_i^z$), it is  useful to  consider $S_z$ basis for its matrix representation. The choice leads to a block diagonal $H$ matrix, with different blocks corresponding to different $S_z^{total}$ values. For even $L$, we focus on the $S_z^{total} = 0$ block, which is the largest block of dimension $M = \frac{L!}{\frac{L}{2}! \, \frac{L}{2}!}$. The diagonalization of the block leads to $M$ non-zero components of an eigenstate $\Psi$ of $H$ with $S_z^{total}=0$; the rest of the components of $\Psi$ are zero (due to $H$ preserving $S_z^{total}$-symmetry). An efficient method to calculate the von Neumann entropy for this case is described in Ref. \cite{zhou2017operator}. With $H$ as a real-symmetric matrix in the chosen basis, the components of its eigenstates  are real variables as well.

We exactly diagonalize the Hamiltonian in eq. \eqref{xxz} for several field strength parameters $h$ and system sizes $L$ while keeping $D$ fixed ($D=1.0$), and, for several values of the $D$ and $h$ for a fixed system size $L=14$. This gives us a large set of eigenstates with two free parameters $h$ and $D$ for many  $L$ values, thus enabling us to explore the $\Lambda$ based analogy among them.  As in the case of QREM, here again we consider both spectral and ensemble averaging, but the analysis is now confined to the neighborhood of the middle of the energy spectrum. For spectral averaging, we choose $100$ eigenstates from the middle of the spectrum (except for $L=12$ where we consider only $50$ eigenpairs), and for the ensemble averaging, the size of the ensemble is chosen based on system size: for $L = 12, 14, 16, 18$, the chosen ensemble sizes are $2000, 1000, 500, 200$ respectively. To proceed further, here again the necessary inputs   $\Delta_e$, $\Omega_e$ and $\chi$ to calculate  $\Lambda$   are determined numerically;  their values for various combinations of $h, D, L$ are displayed in table \ref{tab:chi}.  This, along with substitution of \eqref{yxxz} in eq. (\ref{Lambda}) then gives $\Lambda$ for the $H$ in eq. \eqref{xxz}, also shown in the table \ref{tab:chi}.

\begin{figure}
    \centering
    \includegraphics[width=\textwidth]{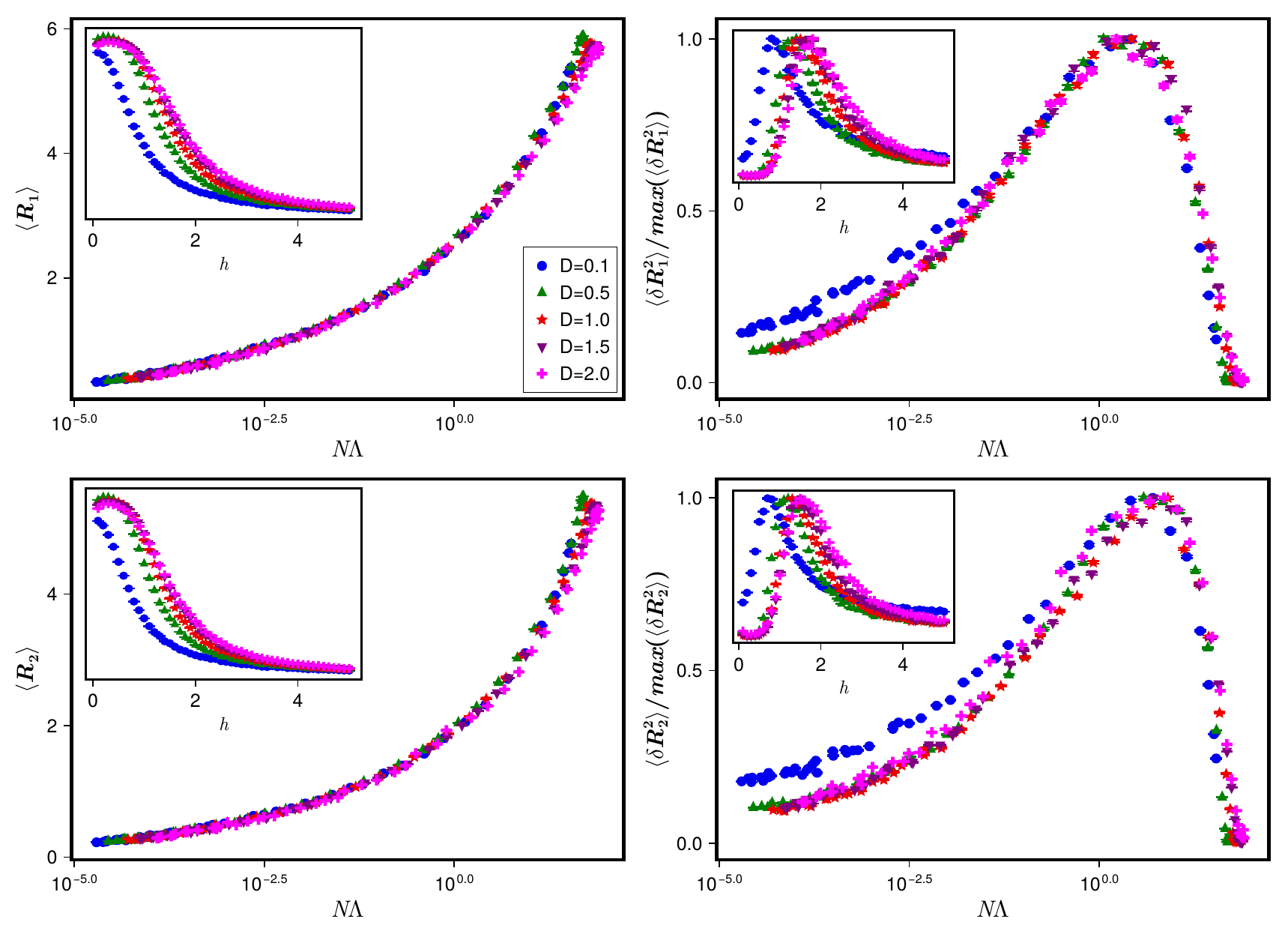}
    \caption{\textbf{Average and variance of von Neumann entropy and second R\'enyi entropy in the RFHM (varying $D$).} The field dynamics of the average and the variance of the (top) von Neumann entropy $R_1$, and (bottom) the second R\'enyi entropy ($R_2$) over various disorder realizations, and $100$ ($50$ for $L=12$) eigenstates per realizations with $h$ varying  for both $D$ and $L$ fixed, is shown. The analysis is displayed for  $D$ fixed at five different values but for $L=14$. As figure indicates, the evolution curves for different $D$ show a common path with $N \Lambda$.}
    \label{fig_r1_r2_delta:rfhm}
\end{figure}

To validate our theoretical claim regarding $\Lambda$ as the single parameter that governs the separability-to-maximum entanglement crossover if the system size is fixed, we study the evolution of entanglement for different anisotropy strength $D$ but fixed system size $L=14$. Figs. \ref{fig_r1_r2_delta:rfhm} show the evolution of the average and variance of $R_1$ and $R_2$ with $\Lambda$ and with $h$ in the inset. As is clear from the figures, the curves for different $D$ indeed collapse onto each other with $\Lambda$, although they show a distinctive evolution with $h$. 

As indicated by many previous studies, increase in $h$ also causes a crossover of the eigenfunction behavior from a localized to ergodic regime. This hints at an underlying connection between the two type of quantum correlations  involved in the separable to entangled state transition and  those in the localization to ergodic state transition. Technically these quantum correlations are of different types, the former measured by entanglement of the local Hilbert space for one subunit of the system with that of another and the latter by the wave-dynamics in whole Hilbert space consisting of all subunits. Indeed, while the entanglement analysis requires at least a bipartite basis, the localization analysis is usually carried out in a monopartite basis. But,  as indicated by our theoretical formulation,  both transitions are essentially analyzed in the same basis i.e. the basis in which the Hamiltonian is represented, a correspondence between two types is indeed expected.

\begin{figure}
    \centering
    \includegraphics[width=\textwidth]{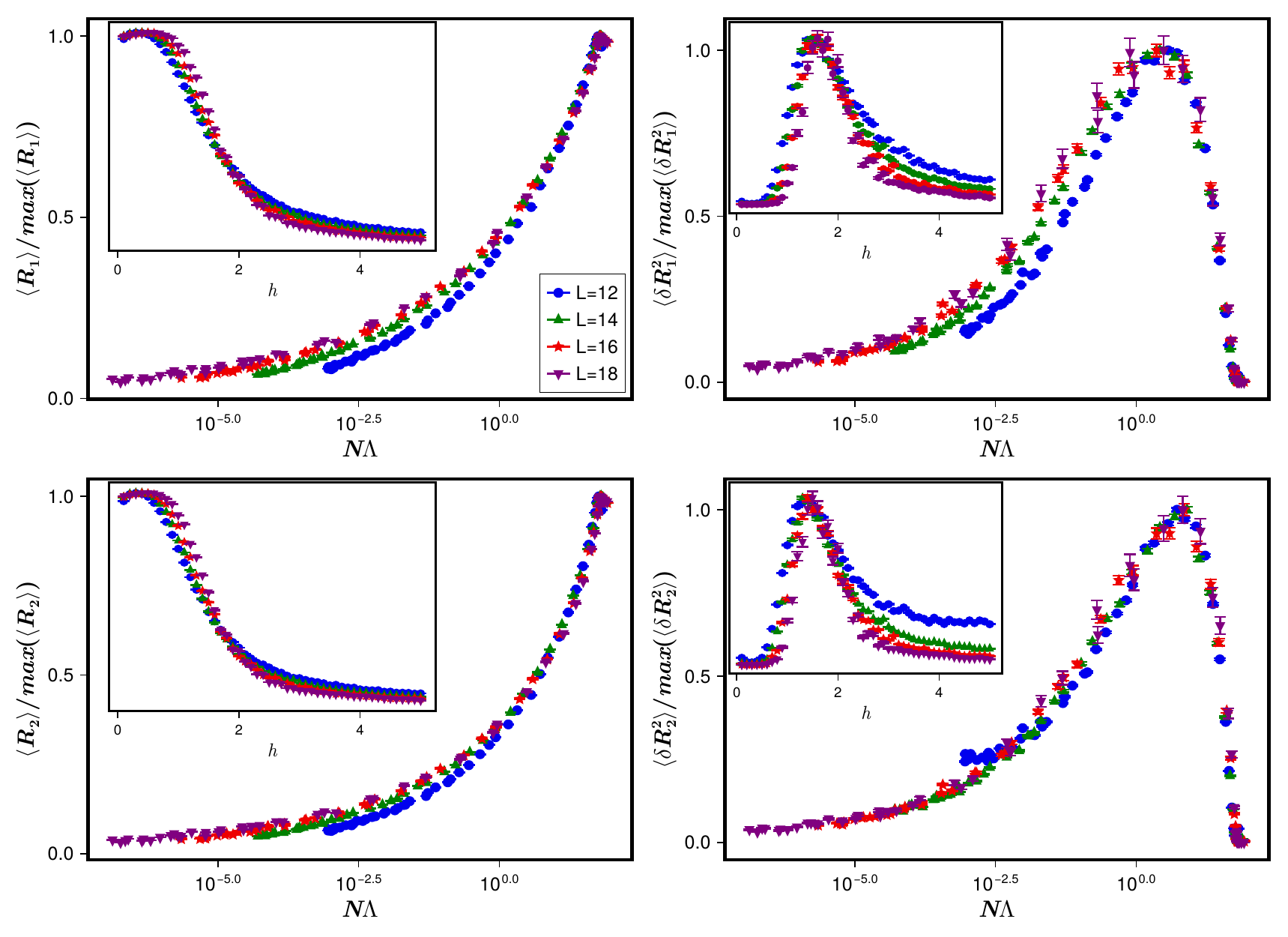}
    \caption{\textbf{Average and variance of von Neumann Entropy and second R\'enyi entropy in the RFHM (varying $L$).} The details here are same as Fig. \ref{fig_r1_r2_delta:rfhm}, however, in this case we study the dynamics for varying system sizes, but fixed  anisotropy parameter $D=1.0$. In this case as well, the evolution curves for different $L$ show a common path with $N \Lambda$. The y-axes have been rescaled by their maximum value to avoid finite-size effects. As can be seen, the evolution curves for different system sizes overlap when plotted with $N \Lambda$, as opposed to with disorder strength $h$ (see inset).}
    \label{fig_r1_r2_size:rfhm}
\end{figure}

The above mentioned connection between the entanglement dynamics and the localization dynamics of a typical eigenfunction give rise to the natural query: whether the former also reveals finite size scaling as well critical behavior (typical of the latter in more than two physical dimensions)? More clearly, based on our theoretical prediction from eq.(\ref{pu}) and eq.(\ref{pdl0}), the eigenfunction statistics depends on both the system size and $\Lambda$, but does this two parameter dependence reduces to a single one in infinite size limit?  To seek the answer, we analyze the size dependence of the entanglement statistics with varying field strength $h$ while keeping $D$ fixed. 
Figs. \ref{fig_r1_r2_size:rfhm} display the $h$ based evolution of the average and variance of $\langle R_1 \rangle$ as well as for  $\langle R_2 \rangle$ for four system sizes $L$ and a fixed $D = 1.0$. The y-axes of the figures are rescaled by their maximum to avoid finite-size effects. As visibly clear from the figures, the evolution curves for different $L$ show an almost collapse for large system sizes if the evolution parameter is chosen as $N \Lambda$ instead of $h$; for a comparison, the inset displays the distinctive behaviour for different $L$ with  $h$ as an evolution parameter. This supports our theoretical claim regarding  $N \Lambda$ as the single  parameter governing the separability to maximum entanglement transition (motivated from the evolution equations, eq. \eqref{avgr0} and eq. \eqref{avgr2}, for instance). 

\begin{figure}
    \centering
    \includegraphics[width=\textwidth]{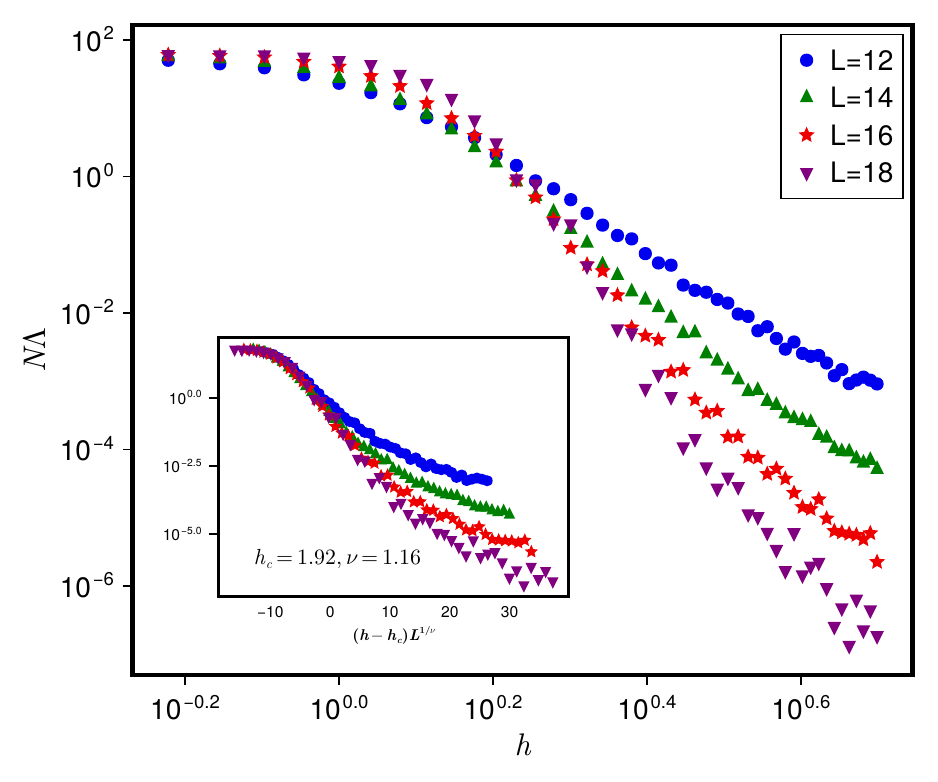}
    \caption{\textbf{Finite-size scaling of $N \Lambda$ in the RFHM.} The figure displays the dependence  of parameter $N \Lambda$ on the field strength $h$ for various system sizes $L$. As clearly visible from the figure, $N \Lambda$ becomes size independent at the critical point $h_c$. A finite-size scaling analysis with the critical point and exponent is also shown in the inset.}
    \label{fig_lambda:rfhm}
\end{figure}

The inset in  Figs. \ref{fig_r1_r2_size:rfhm} also reveals a crossing of the different curves for different system sizes and thus suggesting an underlying phase transition from the separability to entanglement regime. This in turn also suggests the role of $\Lambda$ in characterizing the above phase transition. As mentioned in section \ref{crit}, 
$N \Lambda$  is in general a function of $N$ and approaches, therefore, either $0$ or  $\infty$  in the thermodynamic limit $N \to \infty$, thereby implying  the separable or the maximum entanglement regime of the eigenfunction, respectively  \cite{pseig,psand}.  A subtle conspiracy of the system conditions however may render  $N \Lambda$ independent of $N$; the statistics under such conditions displays a critical behavior and remains different from both the localized or ergodic regimes even in limit $N \to \infty$.  As shown in the Fig. \ref{fig_lambda:rfhm}, this behavior is indeed displayed  for RFHM: $N \Lambda$ for different system sizes with the disorder parameter $h$ intersect each other and hence becomes size-independent at that point, while going to zero and $\infty$ in the localized and the ergodic regimes respectively. A finite-size scaling analysis for $N \Lambda$ is also shown in the inset. The critical point $h_c$ and the critical exponent $\nu$ is different from the values reported in the previous works and is due to the Gaussian randomness considered in this work as opposed to the uniform randomness considered there \cite{luitz2015many,buijsman2019random}. Fig. \ref{fig_lambda:rfhm} then suggests the following

\begin{equation}
    \frac{\bk{R_1}}{\bk{R_1}_{max}} = f(N \Lambda) = f((h-h_c) L^{1/\nu}).
    \label{fig:lambda-fssa}
\end{equation}

An important connection worth emphasizing here is the following: we recall that  the spectral complexity parameter $\Lambda_e$ (defined above eq.(\ref{lamc})) was used in earlier studies \cite{psand} to characterize the localization to delocalization phase transition for the single particle Anderson Hamiltonian. This is also confirmed by our recent studies of the spectral statistics of both QREM and RFHM \cite{shekhar-mbss}. The connection $\Lambda = \chi \Lambda_e$ along with eq.(\ref{fig:lambda-fssa}) again suggests the hidden connection between two types of phase transitions. This also leads to query whether the purely quantum aspect of entanglement as a quantum correlation is only technically connected to other quantum correlations (those with classical limits, e.g., localization and delocalization connected to classical integrability and chaos) or there is something more hidden underneath?

\newpage

\section{Conclusion} \label{conclusion}
In the end we summarize our main ideas, results and open questions. Here we primarily focussed on  the theoretical analysis of  the entanglement dynamics of the many body states as the system conditions vary. As complicated interactions / disorder usually render an exact determination of a many-body Hamiltonian matrix in any physically motivated basis a mathematically intractable task, consideration of their representation by an ensemble is in general unavoidable. This initiated  us  to consider the many-body Hamiltonians which can be well-represented by the multiparametric Gaussian ensembles of Hermitian matrices in a bipartite basis (with uncorrelated / pairwise correlated matrix elements and symmetry resolved to ensure nondegeneracy of the eigenstates). The technical intractability of the original ensemble motivated us to consider a new ensemble representation, referred as the ``complexity ensemble"; the latter is obtained from the former be a mapping of the set of original ensemble parameters  to a new set  leads to a new ensemble representation. The mapping helps because in contrast to original ensemble representation, the ``complexity ensemble" has only one free parameter, basically, a single functional $Y$ of all system conditions  and many invariants, basically, different functionals of basis constants as well as system parameters (former referred in the text as ensemble complexity parameter and the latter as complexity constants). While governed by $Y$, the  dynamics of the new ensemble  in the matrix space is confined to paths constrained  by the complexity constants. (We note the matrix variables remain same for both ensembles, pre and post mapping). This permits us to characterize a  many body physical Hamiltonian (with complicated interactions/ disorder) by a set of complexity constants: the systems with same set of complexity constants evolve along the same path in the matrix space but those with different set evolve along parallel paths. 

The Hamiltonian ensemble in turn gives rise to an infinite range of state ensembles, each representing one of the eigenstates, say $\psi$ and  characterized by the state complexity parameter $\Lambda_{\psi}$ which depends  on energy of the state as well as $Y$ (and thereby system condition). Our analysis indicates the existence of a common evolutionary path of the entanglement measures  for different state ensembles, governed by $\Lambda_{\psi}$. With the system dependence appearing collectively through $\Lambda_{\psi}$, a knowledge of the latter is sufficient to track the effect of varying system conditions on the entanglement measures of a given state. This indicates a potential application of the  $\Lambda_{\psi}$-formulation in achieving the holy grail of the quantum state engineering: the approach to Haar-state starting from an arbitrary quantum state of a many-body Hamiltonian through a controlled variation of the system conditions.  A weak point of our theoretical analysis is lack of an exact formulation for $\chi$, but we have attempted to address the issue by an intuitive guess verified  by a detailed numerical analysis.

Besides Hamiltonian parameters, $\Lambda_{\psi}$ depends on the energy of the eigenstates too and can in general vary for different eigenstates of a given Hamiltonian notwithstanding same set of system conditions. Under a variation of the system conditions however, the entanglement statistics for all eigenstates evolve along a common path lying between separability and maximum entanglement limit. The complexity parameter formulation  can thus be used to compare the relative entanglement characteristics of different eigenstates of a given Hamiltonian under a fixed set of system conditions and can thereby act as a distance measure between two states.  The above statements are however applicable only if the changing system conditions do not change the global constraints class of the Hamiltonian ensemble: as the path depends on the global constraints of the Hamiltonian and thereby on the complexity path constants of the ensemble, it need not remain same if the system conditions affect the global constraints too, e.g., breaking the symmetry conditions, conservation laws too.     

We emphasize that  almost all  theoretical results  in our analysis are derived through exact routes and without approximations. The important claims of our results nonetheless rendered it highly desirable to seek their numerical validation. This however required exact diagonalization of large sized ensembles on many-body systems,  confined not only to a single set of system conditions but also for many such sets to verify the existence of a common complexity path. A numerical determination of $\Lambda_{\psi}$ from various system parameters and eigenvalue-eigenfunction correlations is also a challenging task. While we have achieved the verification based on the numerical analysis of two standard many-body system, namely the quantum random energy model and the random field Heisenberg model, a thorough numerical analysis by  better  numerically equipped research groups is very desirable.  A shortcoming of our current work is a lack of the exact solutions for the $\Lambda$-governed evolution equations for entanglement entropies; each equation however requires solving a complete set of hierarchical equations. 
An alternative route in this context is to calculate the negative moments of the Schmidt eigenvalues from their density of states (instead of pursuing differential equation route) and this work is currently under progress.
We emphasize however that although lack of an exact theoretical formulation for $R_1$ and $R_2$ did not permit us to compare with numerical results in figures, the main theoretical claim about $\Lambda_{\psi}$ as the only parameter governing the statistics is indeed confirmed by our numerics.  Another crucial shortcoming is  a detailed theoretical understanding of $\chi$ is also needed; the latter requires a detailed analysis of the eigenvalues-eigenfunction correlations and their system-dependence and will be pursued elsewhere.  In addition, while the richness of the current analysis has not permitted us to delve on the wealth of potential information contained in complexity constants, this is an important basis for the claimed universality and therefore requires a rigorous  analysis (some examples in this context are discussed in \cite{psds2}).

\section{Acknowledgment}
We thank Dr. Ivan Khaymovich, Nordita for the suggestion to consider quantum random energy model for our numerical analysis. 
We acknowledge National Super computing Mission (NSM) for providing computing resources of `PARAM Shakti' at the IIT Kharagpur, which is implemented by C-DAC and supported by the Ministry of Electronics and Information Technology (MeitY) and Department of Science and Technology (DST), Government of India.  One of the authors (P.S.) is also grateful to SERB, DST, India for the financial support provided for the  research under Matrics grant scheme. D.S. acknowledges financial support from the MHRD through the PMRF scheme.

\bibliographystyle{ieeetr}
\bibliography{references}

@article{zhou2017operator,
  title={Operator entanglement entropy of the time evolution operator in chaotic systems},
  author={Zhou, Tianci and Luitz, David J},
  journal={Physical Review B},
  volume={95},
  number={9},
  pages={094206},
  year={2017},
  publisher={APS}
}

@article{luitz2015many,
  title={Many-body localization edge in the random-field Heisenberg chain},
  author={Luitz, David J and Laflorencie, Nicolas and Alet, Fabien},
  journal={Physical Review B},
  volume={91},
  number={8},
  pages={081103},
  year={2015},
  publisher={APS}
}

@article{nadal2011statistical,
  title={Statistical distribution of quantum entanglement for a random bipartite state},
  author={Nadal, Celine and Majumdar, Satya N and Vergassola, Massimo},
  journal={Journal of Statistical Physics},
  volume={142},
  pages={403--438},
  year={2011},
  publisher={Springer}
}

@article{majumdar2010extreme,
  title={Extreme eigenvalues of Wishart matrices: application to entangled bipartite system},
  author={Majumdar, Satya N},
  journal={arXiv preprint arXiv:1005.4515},
  year={2010}
}

@article{Parolini2020,
   author = {Tommaso Parolini and Gianni Mossi},
   month = {7},
   title = {Multifractal Dynamics of the QREM},
   url = {http://arxiv.org/abs/2007.00315},
   year = {2020}
}

@article{Biroli2021,
   author = {Giulio Biroli and Davide Facoetti and Marco Schiró and Marco Tarzia and Pierpaolo Vivo},
   doi = {10.1103/PhysRevB.103.014204},
   issn = {24699969},
   issue = {1},
   journal = {Physical Review B},
   month = {1},
   publisher = {American Physical Society},
   title = {Out-of-equilibrium phase diagram of the quantum random energy model},
   volume = {103},
   year = {2021}
}

@article{Baldwin2016,
   author = {C. L. Baldwin and C. R. Laumann and A. Pal and A. Scardicchio},
   doi = {10.1103/PhysRevB.93.024202},
   issn = {24699969},
   issue = {2},
   journal = {Physical Review B},
   month = {1},
   publisher = {American Physical Society},
   title = {The many-body localized phase of the quantum random energy model},
   volume = {93},
   year = {2016}
}

@article{Laumann2014,
   author = {C. R. Laumann and A. Pal and A. Scardicchio},
   doi = {10.1103/PhysRevLett.113.200405},
   issn = {10797114},
   issue = {20},
   journal = {Physical Review Letters},
   month = {11},
   publisher = {American Physical Society},
   title = {Many-body mobility edge in a mean-field quantum spin glass},
   volume = {113},
   year = {2014}
}

@article{bianchi2019typical,
  title={Typical entanglement entropy in the presence of a center: Page curve and its variance},
  author={Bianchi, Eugenio and Dona, Pietro},
  journal={Physical Review D},
  volume={100},
  number={10},
  pages={105010},
  year={2019},
  publisher={APS}
}

@article{sup,
title={supplementary material},
author={D. Shekhar and P. Shukla}
}

@article{psds1,
doi = {10.1088/1751-8121/acd9fe},
url = {https://dx.doi.org/10.1088/1751-8121/acd9fe},
year = {2023},
month = {jun},
publisher = {IOP Publishing},
volume = {56},
number = {26},
pages = {265303},
author = {Devanshu Shekhar and Pragya Shukla},
title = {Entanglement dynamics of multi-parametric random states: a single parametric formulation},
journal = {Journal of Physics A: Mathematical and Theoretical}
}

@article{psds2,
  title = {Edge of entanglement in nonergodic states: A complexity parameter formulation},
  author = {Shekhar, Devanshu and Shukla, Pragya},
  journal = {Phys. Rev. E},
  volume = {112},
  issue = {2},
  pages = {024122},
  numpages = {16},
  year = {2025},
  month = {Aug},
  publisher = {American Physical Society},
  doi = {10.1103/txkh-wftd},
  url = {https://link.aps.org/doi/10.1103/txkh-wftd}
}

@article{psds3,
  title = {Distribution of the entanglement entropies of nonergodic quantum states},
  author = {Shekhar, Devanshu and Shukla, Pragya},
  journal = {Phys. Rev. E},
  volume = {112},
  issue = {2},
  pages = {024123},
  numpages = {17},
  year = {2025},
  month = {Aug},
  publisher = {American Physical Society},
  doi = {10.1103/n16c-45rh},
  url = {https://link.aps.org/doi/10.1103/n16c-45rh}
}

@article{pietracaprina2018shift,
  title={Shift-invert diagonalization of large many-body localizing spin chains},
  author={Pietracaprina, Francesca and Mac{\'e}, Nicolas and Luitz, David J and Alet, Fabien},
  journal={SciPost Physics},
  volume={5},
  number={5},
  pages={045},
  year={2018}
}

@Article{slepc,
   author  = "Vicente Hernandez and Jose E. Roman and Vicente Vidal",
   title   = "{SLEPc}: A scalable and flexible toolkit for the solution of eigenvalue problems",
   journal = "ACM Trans. Math. Software",
   volume  = "31",
   number  = "3",
   pages   = "351--362",
   year    = "2005"
}

@article{pseig,
   author = {Pragya Shukla},
   doi = {10.1103/PhysRevE.75.051113},
   issn = {15393755},
   issue = {5},
   journal = {Physical Review E - Statistical, Nonlinear, and Soft Matter Physics},
   month = {5},
   title = {Eigenfunction statistics of complex systems: A common mathematical formulation},
   volume = {75},
   year = {2007},
}

@article{psand,
   author = {Pragya Shukla},
   doi = {10.1088/0953-8984/17/10/020},
   issn = {09538984},
   issue = {10},
   journal = {Journal of Physics Condensed Matter},
   month = {3},
   pages = {1653-1677},
   title = {Level statistics of Anderson model of disordered systems: Connection to Brownian ensembles},
   volume = {17},
   year = {2005},
}

@article{psco,
  title = {Random matrices with correlated elements: A model for disorder with interactions},
  author = {Shukla, Pragya},
  journal = {Phys. Rev. E},
  volume = {71},
  issue = {2},
  pages = {026226},
  numpages = {13},
  year = {2005},
  month = {Feb},
  publisher = {American Physical Society},
  doi = {10.1103/PhysRevE.71.026226},
  url = {https://link.aps.org/doi/10.1103/PhysRevE.71.026226}
}

@article{tmchiral,
   author = {Triparna Mondal and Pragya Shukla},
   doi = {10.1103/PhysRevE.102.032131},
   issn = {24700053},
   issue = {3},
   journal = {Physical Review E},
   month = {9},
   pmid = {33075878},
   publisher = {American Physical Society},
   title = {Spectral statistics of multiparametric Gaussian ensembles with chiral symmetry},
   volume = {102},
   year = {2020},
}

@article{psalt,
  title = {Alternative technique for complex spectra analysis},
  author = {Shukla, Pragya},
  journal = {Phys. Rev. E},
  volume = {62},
  issue = {2},
  pages = {2098--2113},
  numpages = {0},
  year = {2000},
  month = {Aug},
  publisher = {American Physical Society},
  doi = {10.1103/PhysRevE.62.2098},
  url = {https://link.aps.org/doi/10.1103/PhysRevE.62.2098}
}

@article{zyczkowski2011generating,
  title={Generating random density matrices},
  author={{\.Z}yczkowski, Karol and Penson, Karol A and Nechita, Ion and Collins, Benoit},
  journal={Journal of Mathematical Physics},
  volume={52},
  number={6},
  year={2011},
  publisher={AIP Publishing}
}

@article{Collins2016,
   author = {Benoît Collins and Ion Nechita},
   doi = {10.1063/1.4936880},
   issn = {00222488},
   issue = {1},
   journal = {Journal of Mathematical Physics},
   month = {1},
   publisher = {American Institute of Physics Inc.},
   title = {Random matrix techniques in quantum information theory},
   volume = {57},
   year = {2016}
}

@article{luitz-dist,
  title = {Long tail distributions near the many-body localization transition},
  author = {Luitz, David J.},
  journal = {Phys. Rev. B},
  volume = {93},
  issue = {13},
  pages = {134201},
  numpages = {10},
  year = {2016},
  month = {Apr},
  publisher = {American Physical Society},
  doi = {10.1103/PhysRevB.93.134201},
  url = {https://link.aps.org/doi/10.1103/PhysRevB.93.134201}
}

@article{kumar2011entanglement,
  title={Entanglement in random pure states: spectral density and average von Neumann entropy},
  author={Kumar, Santosh and Pandey, Akhilesh},
  journal={Journal of Physics A: Mathematical and Theoretical},
  volume={44},
  number={44},
  pages={445301},
  year={2011},
  publisher={IOP Publishing}
}

@article{page1993average,
  title={Average entropy of a subsystem},
  author={Page, Don N},
  journal={Physical review letters},
  volume={71},
  number={9},
  pages={1291},
  year={1993},
  publisher={APS}
}

@book{bengtsson2017geometry,
  title={Geometry of quantum states: an introduction to quantum entanglement},
  author={Bengtsson, Ingemar and {\.Z}yczkowski, Karol},
  year={2017},
  publisher={Cambridge university press}
}

@article{shekhar2025single,
  title={Single-Particle Entanglement Dynamics in Complex Systems},
  author={Shekhar, Devanshu and Shukla, Pragya},
  journal={Entropy},
  volume={28},
  number={1},
  pages={29},
  year={2025},
  publisher={MDPI},
  doi = {10.3390/e28010029},
  url = {https://doi.org/10.3390/e28010029}
}

@article{buijsman2019random,
  title={Random matrix ensemble for the level statistics of many-body localization},
  author={Buijsman, Wouter and Cheianov, Vadim and Gritsev, Vladimir},
  journal={Physical review letters},
  volume={122},
  number={18},
  pages={180601},
  year={2019},
  publisher={APS}
}

@article{psflat,
  title = {Disorder perturbed flat bands: Level density and inverse participation ratio},
  author = {Shukla, Pragya},
  journal = {Phys. Rev. B},
  volume = {98},
  issue = {5},
  pages = {054206},
  numpages = {14},
  year = {2018},
  month = {Aug},
  publisher = {American Physical Society},
  doi = {10.1103/PhysRevB.98.054206},
  url = {https://link.aps.org/doi/10.1103/PhysRevB.98.054206}
}

@article{psbe,
  title = {Criticality in Brownian ensembles},
  author = {Sadhukhan, Suchetana and Shukla, Pragya},
  journal = {Phys. Rev. E},
  volume = {96},
  issue = {1},
  pages = {012109},
  numpages = {15},
  year = {2017},
  month = {Jul},
  publisher = {American Physical Society},
  doi = {10.1103/PhysRevE.96.012109},
  url = {https://link.aps.org/doi/10.1103/PhysRevE.96.012109}
}

@article{pschiral,
  title = {Statistical analysis of chiral structured ensembles: Role of matrix constraints},
  author = {Mondal, Triparna and Shukla, Pragya},
  journal = {Phys. Rev. E},
  volume = {99},
  issue = {2},
  pages = {022124},
  numpages = {14},
  year = {2019},
  month = {Feb},
  publisher = {American Physical Society},
  doi = {10.1103/PhysRevE.99.022124},
  url = {https://link.aps.org/doi/10.1103/PhysRevE.99.022124}
}

@article{psbe2,
  title={Spectral and strength statistics of chiral Brownian ensemble},
  author={Shukla, Pragya},
  journal={Journal of Physics A: Mathematical and Theoretical},
  volume={54},
  number={27},
  pages={275001},
  year={2021},
  publisher={IOP Publishing}
}

@unpublished{shekhar-mbss,
  author       = {Shekhar, Devanshu and Shukla, Pragya},
  title        = {Spectral statistics of many-body quantum states with evolving system conditions},
  note         = {Manuscript in preparation},
  year         = {2026}
}




\end{document}